\documentclass[sigconf,screen,nonacm,review=false,timestamp=false]{acmart}
\usepackage{popets}

\usepackage{quoting}
\usepackage{hyperref}

\newcommand{\M}{\mathcal{M}}
\newcommand{\data}{\mathbf{x}}
\newcommand{\PL}{\textsf{PrivLoss}}

\newtheorem*{rep@theorem}{\rep@title}
\newcommand{\newreptheorem}[2]{%
\newenvironment{rep#1}[1]{%
 \def\rep@title{#2 \ref{##1}}%
 \begin{rep@theorem}}%
 {\end{rep@theorem}}}
\makeatother

\newreptheorem{theorem}{Theorem}
\newreptheorem{lemma}{Lemma}
\newreptheorem{corollary}{Corollary}
\newreptheorem{example}{Example}

\newenvironment{zquote}[0]
 {%
  \quoting[leftmargin=0cm,rightmargin=0cm, vskip=2mm]%
  \noindent\itshape\ignorespaces
 }
 {\endquoting}

\setcopyright{popets}
\copyrightyear{YYYY}

\acmYear{YYYY}
\acmVolume{YYYY}
\acmNumber{X}
\acmDOI{XXXXXXX.XXXXXXX}
\acmISBN{}
\acmConference{Proceedings on Privacy Enhancing Technologies}
\begin{document}

\title[Interpreting DP in Terms of Disclosure Risk]{Interpreting Differential Privacy in Terms of Disclosure Risk}

%%%%%%%%%%%%%%%% Authors' Info %%%%%%%%%%%%%%%%%

\author{Zeki Kazan}
\affiliation{
  \institution{Duke University}
  \city{Durham}
  \state{North Carolina}
  \country{USA}}
\email{zekican.kazan@duke.edu}
\authornote{Work conducted while interning at TikTok.}

\author{Sagar Sharma}
\affiliation{%
  \institution{TikTok Inc}
  \city{Bellevue}
  \state{Washington}
  \country{USA}}
\email{sagar.sharma@tiktok.com}

\author{Wanrong Zhang}
\affiliation{%
  \institution{TikTok Inc}
  \city{San Jose}
  \state{California}
  \country{USA}}
\email{wanrongzhang@tiktok.com}

\author{Bo Jiang}
\affiliation{%
  \institution{TikTok Inc}
  \city{San Jose}
  \state{California}
  \country{USA}}
\email{bojiang@tiktok.com}

\author{Qiang Yan}
\affiliation{%
  \institution{TikTok Inc}
  \city{Shanghai}
  \country{China}}
\email{yanqiang.mr@tiktok.com}


\renewcommand{\shortauthors}{Kazan et al.}

\begin{abstract}
    As the use of differential privacy (DP) becomes widespread, the development of effective tools for reasoning about the privacy guarantee becomes increasingly critical. In pursuit of this goal, we demonstrate novel relationships between DP and measures of statistical disclosure risk. We suggest how experts and non-experts can use these results to explain the DP guarantee, interpret DP composition theorems, select and justify privacy parameters, and identify worst-case adversary prior probabilities.
\end{abstract}


\maketitle

\section{Introduction} \label{sec:intro}

Over recent years, differential privacy (DP) has become increasingly prominent, with many large companies \cite{aktay2020google, herdaugdelen2020protecting, tang2017privacy}, government agencies \cite{abowd20222020, TumultLabs_DeptEd}, and others \cite{Spectus, adeleye2023publishing} turning to DP for the collection, analysis, and release of quantities involving sensitive data.
DP provides a strong privacy guarantee \cite{dwork2006calibrating}, has desirable properties (e.g., composition \cite{steinke2022composition}, post-processing immunity \cite{dwork2014algorithmic}),
and can be straightforward to implement via existing tools, such as OpenDP 
\cite{gaboardi2020programming}. 

But the use of DP creates challenges. 
It can be difficult to interpret what the use of a given privacy budget %implies about the effect of an individual's inclusion in the underlying database on their disclosure risk.
% implies about the risk to an individual in the data's privacy.
implies about the privacy risk for individuals in the underlying data.
Because of this, selecting and justifying an appropriate privacy budget can be challenging, as discussed in prior work, e.g., \cite{kazan2024prior, lee2011much}. 
Additionally, explanations and interpretations of the privacy protection from DP often have to be tailored to the particular audience. For example, a layman may be interested in determining if the privacy protection is sufficient before participating in a data collection scheme, a regulator may be interested in %understanding the chances of re-identification of an individual in a dataset, 
verifying that an organization has adequately addressed the potential for privacy risks,
while a data curator may be interested in selecting a sufficient privacy budget to meet regulators' expectations. 

To address these challenges, we present comprehensive Bayesian semantic characterizations of pure, approximate, and probabilistic DP. Our primary contribution is a series of results relating these privacy definitions to disclosure risk criteria, such as bounds on an adversary's posterior probability, posterior-to-prior ratio, and posterior-to-prior difference. Prior work \cite{dwork2006calibrating, lee2011much, kazan2024prior, kifer2014pufferfish, wood2018differential} focuses on semantics of pure DP; as far as we are aware, our results are the first to characterize approximate DP---the notion most common in practice---in this manner. We also demonstrate equivalences between probabilistic DP and posterior bounds as well as between pure DP and bounds on a posterior-to-prior difference, which we believe to be novel.
%Of particular note are the results for approximate DP, which is the notion most common in practice, but which has received less attention than pure DP, where the semantics were studied in, e.g., \cite{dwork2006calibrating, lee2011much, kazan2024prior, kifer2014pufferfish, wood2018differential}. As far as we are aware, our results are the first to characterize approximate DP in this manner as well as the first to demonstrate equivalence between pure DP and bounds on an adversary's posterior-to-prior difference. 
%In this work, we present a novel Bayesian semantic characterization of approximate DP. Prior work has generally focused on Bayesian semantics of pure DP \cite{dwork2006calibrating, lee2011much, kazan2024prior, kifer2014pufferfish, wood2018differential}, while approximate DP---the notion more common in practice---has received comparably little attention \cite{kasiviswanathan2014semantics}. Our primary contribution is a result demonstrating equivalence between approximate DP and a bound on an adversary's posterior probability that a target is in the underlying database used for a data release. As far as we are aware, this result is the first to demonstrate an equivalence between approximate DP and bounds on Bayesian adversaries' posterior probabilities. 
Finally, we outline a number of potential applications of these results:
\begin{enumerate}
    \item Explaining aspects of DP to interested parties with varying technical backgrounds. 
    \item Providing an alternate, intuitive perspective on popular DP composition theorems.
    %Interpreting the disclosure risk from the composition of many DP mechanisms, such as daily DP releases or sequential queries of a DP database.
    \item Extending existing methods \cite{kazan2024prior} for selecting the privacy budget of a DP release from pure to approximate DP. 
    %Extending existing methods \cite{kazan2024prior} for selecting the privacy budget of a DP release by enforcing bounds on an adversary's posterior distribution from pure to approximate DP.
    \item Identifying an adversary's worst-case prior probability with respect to a given disclosure risk criteria.
    %Identifying the worst-case prior probability an adversary may have with respect to a given criteria, e.g., the adversary's posterior-to-prior ratio or difference, on which a data curator may focus an analysis. 
\end{enumerate}

To briefly demonstrate, consider the following example, adapted from Nanayakkara et al.~(2023) \cite{nanayakkara2023chances}.

\begin{example} \label{ex:simple}
    Employees are required to evaluate their manager, including answering the \texttt{YES}/\texttt{NO} question \textit{``Do you feel adequately supported by your manager?''} Managers will receive a summary with the count of each response. There is concern that if an employee believes their manager knows all others will respond \texttt{YES}, they risk retaliation should they respond \texttt{NO}. To avoid this scenario, noise is added to summaries before they are released to the managers. The employees and managers are not provided with the details of the noise mechanism, but are told it satisfies ($\varepsilon = 0.1, \delta = 10^{-7})$-DP.
\end{example}

Suppose that, before viewing the summary counts, the manager believes there is a 50\% probability a particular employee will respond \texttt{NO}. If the release had satisfied pure $(\varepsilon = 0.1, \delta = 0)$-DP, existing results (see Section \ref{sec:prior_work}) would guarantee that the probability the manager assigns to the employee responding \texttt{NO} will be between 48\% and 52\% after observing the summary counts. It would be ideal if, for $(\varepsilon = 0.1, \delta = 10^{-7})$-DP, we could state that this guarantee holds with high probability. Unfortunately, approximate DP is not equivalent to pure DP holding with probability $1-\delta$. We show, however, that with at least 99\% probability, the manager's final probability will indeed be between 48\% and 52\%.

%To interpret the privacy guarantee via our results, suppose that, before viewing the summary counts, the manager believes there is a 50\% probability a particular employee will respond \texttt{NO}. If indeed the manager were to know all other employees responded \texttt{YES}, then with 99\% probability, the probability the manager assigns to the employee responding \texttt{NO} will be between 48\% and 52\% after observing the summary counts. In Section \ref{sec:understand_DP}, we return to this example and describe how alternate bounds can be provided without making an assumption about the manager's initial probability as well as how the company might describe the privacy guarantee when asking employees to provide their responses.

The remainder of this work is structured as follows. In Section \ref{sec:background}, we review relevant results in the DP literature and provide an overview of prior work on Bayesian semantics of DP. In Section \ref{sec:results}, we outline the setting and state our main theoretical results. In Section \ref{sec:applications}, we discuss several practical applications of our results. In Section \ref{sec:discuss}, we provide concluding thoughts and discuss avenues for future work. Finally, in %Proofs of all results are provided in 
Appendix \ref{sec:proofs}, we provide proofs of all results.

\section{Background and Related Work} \label{sec:background}

In this section, we review the definitions of DP and several of its variants as well as key foundational results. We then summarize prior work on Bayesian semantics of DP and its variants.

\subsection{Differential Privacy} \label{sec:DP}

We begin by reviewing the definition of DP. Databases $\data, \data'$ are said to be \textit{neighboring} if they differ in the presence or absence of one observation. What an observation represents is referred to as the \textit{privacy unit}.\footnote{In real-world deployments, the privacy unit is often either one user, e.g., \cite{abowd20222020, pereira2021us}, or one user per day, e.g., \cite{adeleye2023publishing, aktay2020google, herdaugdelen2020protecting}. Other privacy units, however, are sometimes preferred, e.g., an individual's action \cite{DVN/TDOAPG_2020} or whether a user made a trip in a given week \cite{bassolas2022reply}.}
A mechanism $M$ satisfies DP if its output does not differ substantially between any neighboring databases. This is formalized as follows.

\begin{definition} 
    \cite{dwork2006our}
    A mechanism $M$ satisfies $(\varepsilon, \delta)$-differential privacy if for any $S \in \textsf{Range}(M)$ and for any pair of neighboring databases, $\data$ and $\data'$, $P[M(\data) \in S] \leq e^{\varepsilon} P[M(\data') \in S] + \delta.$
    % \begin{align*}
    %     P[M(\data) \in S] \leq e^{\varepsilon} P[M(\data') \in S] + \delta.
    % \end{align*}
\end{definition}

If $\delta = 0$, $M$ is said to satisfy \textit{pure DP}, while if $\delta > 0$, $M$ is said to satisfy \textit{approximate DP}. An alternative way of defining DP involves considering the privacy loss random variable (PLRV), as defined in \cite{dwork2014algorithmic, steinke2022composition}.

\begin{definition}
    Let $f_{P \Vert Q}(y) = \log\left(P(y)/Q(y)\right)$ for probability distributions $P$ and $Q$.
    % \begin{align*}
    %     f_{P \Vert Q}(y) = \log\left(\frac{P(y)}{Q(y)}\right).
    % \end{align*}
    Then the privacy loss random variable is given by $Z = f_{P \Vert Q}(Y)$ for $Y \leftarrow P$ and is denoted $\PL(P \Vert Q)$.
\end{definition}

Pure DP is equivalent to bounding a PLRV, as follows.

\begin{theorem} \label{thm:pure_DP_PLRV}
    \cite{steinke2022composition}
    $M$ satisfies $(\varepsilon, 0)$-DP if and only if for any neighboring databases $\data$ and $\data'$, $P[-\varepsilon \leq Z \leq \varepsilon] = 1$, where
    $Z = \PL(M(\data) \Vert M(\data'))$.
\end{theorem}

Approximate DP mechanisms are sometimes incorrectly characterized as ``mechanisms that achieve pure DP with probability $1-\delta$.'' This characterization gives rise to an alternate definition: probabilistic differential privacy (PDP).

\begin{definition} 
    \cite{machanavajjhala2008privacy}
    $M$ satisfies $(\varepsilon, \delta)$-probabilistic differential privacy if for any neighboring databases $\data$ and $\data'$, $P[-\varepsilon \leq Z \leq \varepsilon] \geq 1-\delta$, where
    $Z = \PL(M(\data) \Vert M(\data'))$.
\end{definition}

PDP is rarely used in practice, since it is not preserved under post-processing \cite{kifer2012axiomatic}. Notably, mechanisms that satisfy $(\varepsilon, \delta)$-DP need not satisfy $(\varepsilon, \delta)$-PDP when $\delta > 0$; see \cite{meiser2018approximate} for a counterexample. The converse, however, does hold:

%However, PDP implies approximate DP, which does not have this drawback.

\begin{theorem} \label{thm:PDP_to_DP}
    \cite{dwork2014algorithmic, zhao2019reviewing} If $M$ satisfies $(\varepsilon, \delta)$-PDP, then $M$ satisfies $(\varepsilon, \delta)$-DP.
\end{theorem}

%The converse of this result does not hold for $\delta > 0$; see \cite{meiser2018approximate} for a counterexample. 
Next, we examine zero-concentrated differential privacy (zCDP), a commonly used variant of DP which can be defined in terms of a PLRV.

\begin{definition}
    \cite{bun2016concentrated} 
    $M$ satisfies $\rho$-zCDP if for any $\alpha > 1$ and any pair of neighboring databases, $\data$ and $\data'$, $Z = \PL(M(\data) \Vert M(\data'))$ satisfies $E\left[e^{(\alpha-1)Z}\right] \leq e^{\alpha(\alpha-1)\rho}$.
    % \begin{align*}
    %     E\left[e^{(\alpha-1)Z}\right] \leq e^{\alpha(\alpha-1)\rho}.
    % \end{align*}
\end{definition}

If $M$ satisfies $\rho$-zCDP, then for any $\delta > 0$, it satisfies $(\varepsilon, \delta)$-DP .

\begin{theorem} \label{thm:zCDP_to_approx_DP}
    \cite{bun2016concentrated} If $M$ satisfies $\rho$-zCDP, then for any $\delta > 0$, $M$ satisfies $(\varepsilon, \delta)$-DP for $\varepsilon = \rho + 2\sqrt{\rho\log(1/\delta)}$.
\end{theorem}

An important property of DP is composition. A few important composition results are summarized below.

\begin{theorem} \label{thm:DP_comp}
    Let $M_j$ satisfy $(\varepsilon_j, \delta_j)$-DP for $j \in \{1, \ldots, k\}$. Define $M = (M_1, \ldots, M_k)$. Then,
    \begin{enumerate}
        \item \textbf{Basic Composition \cite{dwork2006our}:} $M$ satisfies $(\varepsilon, \delta)$-DP for $\varepsilon = \sum_j \varepsilon_j$ and $\delta = \sum_j \delta_j$.

        \item \textbf{Advanced Composition \cite{dwork2010boosting}:} $M$ satisfies $(\varepsilon, \delta)$-DP for $\delta > \sum \delta_j$ and
        \begin{align*}
            \varepsilon =\sum_{j=1}^k \varepsilon_j(e^{\varepsilon_j} - 1) + \sqrt{2 \sum_{j=1}^k\varepsilon_j^2 \log\left(\frac{1}{\delta-\sum\delta_j}\right)}.
        \end{align*}

        \item \textbf{Optimal Composition\footnote{We focus on homogeneous optimal composition. \cite{murtagh2015complexity} demonstrate that the heterogeneous case---where the $\varepsilon_j$ differ---is $\#P$-complete.} \cite{kairouz2015composition}:} If $\varepsilon_j = \varepsilon'$ and $\delta_j = \delta'$ for all $j$, then for any $\ell \in \{0, 1, \ldots, \lfloor k/2 \rfloor\}$, $M$ satisfies $(\varepsilon, \delta)$-DP for $\varepsilon = (k - 2\ell)\varepsilon'$, $\delta = 1 - (1-\delta')^k(1-\delta_\ell)$, and 
        \begin{align*}
            \delta_\ell = \sum_{j=0}^{\ell - 1} \binom{k}{j}\left(e^{(k-j)\varepsilon'} - e^{(k - 2\ell + j)\varepsilon'}\right)/\left(1+e^{\varepsilon'}\right)^k.
            % \delta_\ell = \sum_{j=0}^{\ell - 1} \binom{k}{j}\frac{\left(e^{(k-j)\varepsilon'} - e^{(k - 2\ell + j)\varepsilon'}\right)}{\left(1+e^{\varepsilon'}\right)^k}.
        \end{align*}
    \end{enumerate}
\end{theorem}

Finally, zCDP too satisfies a composition theorem.

\begin{theorem}
    \cite{bun2016concentrated}
    %Let $M_1, \ldots, M_k$ be such that $M_j$ satisfies $\rho_j$-zCDP. Define $M = (M_1, \ldots, M_k)$. Then, $M$ satisfies $\rho$-zCDP for $\rho = \sum_j \rho_j$.
    Let $M_j$ satisfy $\rho_j$-zCDP for $j \in \{1, \ldots, k\}$. Define $M = (M_1, \ldots, M_k)$. Then, $M$ satisfies $(\sum_j \rho_j)$-zCDP. 
\end{theorem}

\subsection{Bayesian Semantics of Differential Privacy} \label{sec:prior_work}

The issue of generalizable knowledge complicates Bayesian semantics of DP.
For example, if the relationship between smoking and lung cancer is unknown, a study uncovering evidence that smoking causes lung cancer can lead to a large change in an adversary's belief about whether a person is at elevated risk of lung cancer, whether they participated in the study or not \cite{dwork2013toward, kasiviswanathan2014semantics}. Without additional assumptions, we cannot disentangle this population-level information from individual-level quantities, such as an adversary's gain in knowledge due to a person's inclusion. There are three methods in the literature for ensuring an adversary's change in beliefs is only due to individual-level quantities.

The first method involves considering a ``strong adversary'' possessing the complete information of all individuals in the data except their target. In the paper proposing DP, Dwork et al.~(2006) \cite{dwork2006calibrating} demonstrates that pure DP is equivalent to bounds on a strong adversary's posterior-to-prior ratio for binary mechanisms\footnote{Each work in this section considers different events for the adversary's prior and posterior probabilities. E.g., \cite{dwork2006calibrating} examines the event that an arbitrary binary predicate returns $1$. For clarity, we do not detail the events considered by each work; interested readers should refer to the cited works.}. Later works show that pure DP implies bounds on a strong adversary's posterior-to-prior odds ratio \cite{abowd2008protective} and bounds on a strong adversary with uniform prior distribution's posterior probabilities \cite{lee2011much, john2021decision, pankova2022interpreting}. 
It has also been shown that zCDP implies a bound on a strong adversary's posterior-to-prior ratio \cite{kifer2022bayesian}. Other works use simulation studies to examine a strong adversary’s realized posterior beliefs under both pure DP \cite{mcclure2012differential} and zCDP discrete Gaussian noise infusion \cite{kazan2025assessing}.
%\cite{mcclure2012differential} examines a strong adversary's realized posterior beliefs in a simulation study. For zCDP, \cite{kifer2022bayesian} show that zCDP implies a bound on a strong adversary's posterior-to-prior ratio and \cite{kazan2025assessing} examine the exact form of a strong adversary's posterior for releases via zCDP discrete Gaussian noise infusion \cite{canonne2020discrete}.

The second method involves assuming the presence or record value of each individual in a dataset is independent of those of other individuals. This is a strong assumption, but is analogous to common assumptions in classical statistics. Under this assumption,
it has been shown that pure DP implies a bound on an adversary's posterior-to-prior ratio \cite{kazan2024prior} and is equivalent to bounds on an adversary's posterior-to-prior odds ratio \cite{kifer2014pufferfish}.
% \cite{kifer2014pufferfish} demonstrates equivalence between pure DP and bounds on an adversary's posterior-to-prior odds ratio and \cite{kazan2024prior} shows that pure DP implies a bound on an adversary's posterior-to-prior ratio.

The third method involves comparing an adversary's posterior distribution where the target is present to the counterfactual where the target is not present. It has been shown that pure DP implies bounds on both the posterior as a function of the counterfactual posterior \cite{wood2018differential, wood2020designing} and the ratio of the posterior to a posterior with the target's attributes replaced by a draw from the counterfactual posterior \cite{kifer2022bayesian}. It has also been shown that approximate DP implies a bound on the total variation distance between the posterior distribution and the counterfactual posterior distribution \cite{kasiviswanathan2014semantics}.
%For approximate DP, \cite{kasiviswanathan2014semantics} demonstrates a bound on the total variation distance between the posterior and counterfactual posterior distributions. For pure DP, \cite{wood2018differential, wood2020designing} references bounds on the posterior as a function of the counterfactual posterior and \cite{kifer2022bayesian} demonstrates bounds on the ratio of the posterior to a posterior with the target's attributes replaced by a draw from the counterfactual posterior.

Cummings et al.~(2024) \cite{cummings2024attaxonomy} recently proposed the ``Attaxonomy'' framework for more explicitly categorizing the adversarial assumptions described above; we direct interested readers to that work for more details.
Generally, no set of assumptions 
%none of these methods
is preferred over the others, but one may make more sense for a given application. For example, it may not be reasonable to assume observations are independent for a sensitive dataset involving the disease status of individuals with a contagious disease. For this work, we consider the first set of assumptions involving a strong adversary, since we believe it is most effective for crafting intuitive explanations about the DP guarantee. We hypothesize in Section \ref{sec:discuss} that analogous results to those presented in Section \ref{sec:results} may hold under the other assumptions.

\section{Theoretical Results} \label{sec:results}

We now present our main results. We first outline our membership inference attack setting and then demonstrate relationships between disclosure risk metrics and each of pure, approximate, and probabilistic DP in this setting. 
%DP is equivalent to bounds on the distribution of an adversary's posterior probability. We then examine the implications for adversaries' posterior-to-prior ratios and differences. Full proofs of all results are in Appendix \ref{sec:proofs}.

\subsection{The Membership Inference Attack Setting} \label{sec:setting}

We consider the following membership inference attack setting. An organization possesses a potentially sensitive database $\data$ and releases the output of a mechanism $M(\data)$. An adversary uses $M(\data)$ to infer whether a target, indexed by $i$, is present in $\data$. Let $\M$ denote the adversary's model based on their prior information and let $I_i$ be an indicator variable for the presence of the target in $\data$. If $M$ is the composition of multiple mechanisms, $\data$ is assumed to be the same across all mechanisms.

%We take the first of the adversarial assumptions discussed in Section \ref{sec:prior_work}. 
Following \cite{lee2011much}, we protect against a worst-case scenario by considering a knowledgeable adversary that possesses $\data_{-i}$, the complete information of all individuals in $\data$ except the target. This corresponds to the first method discussed in Section \ref{sec:prior_work}. We also assume the adversary knows the target's attributes; this information is incorporated into $\M$. All the adversary does not know is whether $i$ is present in $\data$. Following \cite{kazan2024prior}, we define %the adversary's prior and posterior probabilities as follows. Let 
$p_i = P_\M[I_i = 1]$ to be the adversary's prior probability $i$ is in $\data$. We let $Y$ be a random variable representing the output of $M$; the adversary is unsure whether $Y \leftarrow M(\data)$ or $Y \leftarrow M(\data_{-i})$. %the adversary knows that either $Y \leftarrow M(\data)$ or $Y \leftarrow M(\data_{-i})$. 
We define the random variable $X_i = f_i(Y, p_i)$ to represent the distribution of the adversary's posterior beliefs, where for an observed outcome $y$, the adversary's posterior probability is given by
\begin{align} \label{eq:f}
    f_i(y, p_i) = P_{\M}[I_i = 1 \mid Y = y].
\end{align}
% For an observed outcome $y$, the adversary's posterior is given by the function
% \begin{align} \label{eq:f}
%     f_i(y, p_i) = P_{\M}[I_i = 1 \mid Y = y].
% \end{align}
% We define the random variable $X_i = f_i(Y, p_i)$ to represent the distribution of the adversary's posterior beliefs.

\subsection{Results for Probabilistic DP} \label{sec:results_PDP}

We first derive Bayesian semantics for PDP, which we will leverage in the following sections to attain corresponding results for pure and approximate DP. We begin with a novel result equating $X_i$ to functions of PLRVs, which will be an important tool in the proofs of the results that follow. See Appendix \ref{proof:PLRV_posterior} for a full proof.

%We will now demonstrate equivalence between PDP and a bound on the distribution of $X_i$. To begin, we prove a novel result equating $X_i$ to functions of privacy loss random variables.

\begin{theorem} \label{thm:PLRV_posterior}
    For the setting described in Section \ref{sec:setting}, define the privacy loss random variables $Z_i = \PL(M(\data) ~\Vert~ M(\data_{-i}))$ and $Z_i' = \PL(M(\data_{-i}) ~\Vert~ M(\data))$. Then 
    \begin{align} \label{eq:posterior_PLRV}
        X_i = 
        \begin{cases}
            \frac{p_i}{p_i + (1-p_i)e^{-Z_i}}, & \mbox{if } Y \leftarrow M(\data); \\
            \frac{p_i}{p_i + (1-p_i)e^{Z_i'}}, & \mbox{if } Y \leftarrow M(\data_{-i}).
        \end{cases}
    \end{align}
    % if $Y \leftarrow M(\data)$,
    % \begin{align} \label{eq:posterior_PLRV_1}
    %     X_i = \frac{p_i}{p_i + (1-p_i)e^{-Z_i}}.
    % \end{align}
    % If $Y \leftarrow M(\data_{-i})$,
    % \begin{align} \label{eq:posterior_PLRV_2}
    %     X_i = \frac{p_i}{p_i + (1-p_i)e^{Z_i'}}.
    % \end{align}
\end{theorem}

For fixed $p_i$, the relationships between $X_i$ and the PLRVs in (\ref{eq:posterior_PLRV}) are one-to-one functions. Thus, bounds on $Z_i$ and $Z_i'$ imply bounds on $X_i$. Intuitively then, since PDP is defined in terms of bounds on PLRVs, PDP is equivalent to a bound on $X_i$ holding for any $\data$ and all potential targets $i$. Theorem \ref{thm:PDP_posterior}---proved in Appendix \ref{proof:PDP_posterior}---formalizes this intuition.

\begin{theorem} \label{thm:PDP_posterior}
    $M$ satisfies $(\varepsilon, \delta)$-PDP if and only if in the setting described in Section \ref{sec:setting}, for any database $\data$, target $i$, $p_i \in [0,1]$, and under both $Y \leftarrow M(\data)$ and $Y \leftarrow M(\data_{-i})$, the distribution of $X_i$ satisfies
    \begin{align} \label{eq:eps_delta_post}
        P\left[\frac{p_i}{p_i + (1-p_i)e^{\varepsilon}} \leq X_i \leq \frac{p_i}{p_i + (1-p_i)e^{-\varepsilon}} \right] \geq 1-\delta.
    \end{align}
\end{theorem}

The bounds in (\ref{eq:eps_delta_post}) are similar to bounds that appear in prior work in the context of pure DP \cite{kazan2024prior, wood2018differential}. These prior results, however, demonstrate only the one-way implication that DP implies a bound involving an adversary's posterior probability. As far as we are aware, our work is both the first to prove the reverse implication---that bounds on adversaries' posterior probabilities imply PDP---and the first to prove a result for $\delta > 0$ for bounds of the form in (\ref{eq:eps_delta_post}). Of course, since PDP is rarely used in practice, this result is of mainly theoretical interest. Combining this result with results relating pure and approximate DP to PDP yield results of more practical interest, as we discuss in Sections \ref{sec:results_pure_DP} and \ref{sec:results_approx_DP}.

The prior work discussed in Section \ref{sec:prior_work} mainly considers notions of relative disclosure risk, often the ratio of an adversary's posterior probability to their prior probability. We now demonstrate that PDP is equivalent to a prior-independent bound on this ratio. See Appendix \ref{proof:PDP_ratio} for a full proof.

\begin{theorem} \label{thm:PDP_ratio}
    For the setting described in Section \ref{sec:setting}, let the posterior-to-prior ratio be given by
    \begin{align} \label{eq:r}
        r_i(y, p_i) = \frac{P_{\M}[I_i = 1 \mid Y = y]}{P_{\M}[I_i = 1]}.
    \end{align}
    Let $R_i = r_i(Y, p_i)$. Then, $M$ satisfies $(\varepsilon, \delta)$-PDP if and only if for any database $\data$, target $i$, $p_i \in (0,1]$, and under both $Y \leftarrow M(\data)$ and $Y \leftarrow M(\data_{-i})$,
    \begin{align} \label{eq:eps_delta_ratio}
        P\left[e^{-\varepsilon} \leq R_i \leq e^{\varepsilon} \right] \geq 1-\delta.
    \end{align}
\end{theorem}

Some critics of DP argue that the well-established relationships between variants of DP and measures of relative risk imply that these privacy definitions are based on an ``inappropriate measure of disclosure risk'' \cite{hotz2022balancing}. The authors of \cite{hotz2022balancing} argue that an individual in a database should be most concerned about absolute measures of risk
since ``the individual will care a great deal about small increases in the disclosure risk if the probability of disclosure is already high, but may not be bothered by even a large relative increase in risk from data release if the probability of disclosure remains low in absolute terms after release.'' We now show that bounds on the posterior-to-prior difference are similarly related to PDP. Lemma \ref{lem:diff_worst}--proved in Appendix \ref{proof:diff_worst}---is important for analyzing the properties of the posterior-to-prior difference.

\begin{lemma} \label{lem:diff_worst}
    Suppose that for all $p_i \in [0,1]$, a random variable $D_i$ is such that
    \begin{align} \label{eq:D_i_bounds}
        P\left[\frac{p_i}{p_i + (1-p_i)e^{\varepsilon}} - p_i \leq D_i \leq \frac{p_i}{p_i + (1-p_i)e^{-\varepsilon}} - p_i \right] \geq 1-\delta.
    \end{align}
    Then,
    \begin{enumerate}
        \item The lower limit of the interval in (\ref{eq:D_i_bounds}) is minimized when $p_i = 1/ (1+e^{-\varepsilon/2})$.
    
        \item The upper limit of the interval in (\ref{eq:D_i_bounds}) is maximized when $p_i = 1/(1+e^{\varepsilon/2})$.
    \end{enumerate}
\end{lemma}

We leverage Lemma \ref{lem:diff_worst} to prove the following result, whereby a mechanism satisfying probabilistic DP implies a prior-independent bound on an adversary's posterior-to-prior difference. See Appendix \ref{proof:PDP_diff} for a full proof.

\begin{theorem} \label{thm:PDP_diff}
    For the setting described in Section \ref{sec:setting}, let the posterior-to-prior difference be given by
    \begin{align} \label{eq:d}
        d_i(y, p_i) = P_{\M}[I_i = 1 \mid Y = y] - P_{\M}[I_i = 1].
    \end{align}
    Let $D_i = d_i(Y, p_i)$. Then,
    \begin{enumerate}
        \item If $M$ satisfies $(\varepsilon, \delta)$-PDP, then for any database $\data$, target $i$, $p_i \in [0,1]$, and under both $Y \leftarrow M(\data)$ and $Y \leftarrow M(\data_{-i})$,
        \begin{align} \label{eq:eps_delta_diff}
            P\left[-\frac{e^{\varepsilon/2} - 1}{e^{\varepsilon/2} + 1} \leq D_i \leq \frac{e^{\varepsilon/2} - 1}{e^{\varepsilon/2} + 1} \right] \geq 1-\delta.
        \end{align}

        \item If $M$ is such that for any database $\data$, target $i$, $p_i \in [0,1]$, and under both $Y \leftarrow M(\data)$ and $Y \leftarrow M(\data_{-i})$,
        \begin{align} \label{eq:eps_delta_diff2}
            P\left[-\frac{e^{\varepsilon/2} - 1}{e^{\varepsilon/2} + 1} \leq D_i \leq \frac{e^{\varepsilon/2} - 1}{e^{\varepsilon/2} + 1} \right] \geq 1-\delta,
        \end{align}
        then $M$ satisfies both $(\varepsilon, 2\delta)$-PDP and $(\tilde{\varepsilon}, \delta)$-PDP for $\tilde{\varepsilon} = \log(3e^{\varepsilon/2} - 1) - \log(3 - e^{\varepsilon/2})$.
    \end{enumerate}
\end{theorem}

Notably, unlike for the posterior-to-prior ratio, there is a loss in parameters for one direction of the result. We provide a simple example in Appendix \ref{proof:counterexample} demonstrating that the converse to part (1) of Theorem \ref{thm:PDP_diff} does not hold.

\subsection{Results for Pure DP} \label{sec:results_pure_DP}

Theorem \ref{thm:pure_DP_PLRV}, which states that $(\varepsilon, 0)$-PDP is equivalent to $\varepsilon$-DP, implies a series of corollaries to our main results. These corollaries are proved in Appendix \ref{proof:pure_DP}. 
First, Theorem \ref{thm:PDP_posterior} implies that DP is equivalent to a bound on an adversary's posterior probability holding with probability $1$.

\begin{corollary} \label{cor:pure_DP_posterior}
    $M$ satisfies $(\varepsilon, 0)$-DP if and only if in the setting described in Section \ref{sec:setting}, for any database $\data$, target $i$, $p_i \in [0,1]$, and under both $Y \leftarrow M(\data)$ and $Y \leftarrow M(\data_{-i})$, the distribution of $X_i$ satisfies
    \begin{align} \label{eq:eps_post}
        P\left[\frac{p_i}{p_i + (1-p_i)e^{\varepsilon}} \leq X_i \leq \frac{p_i}{p_i + (1-p_i)e^{-\varepsilon}} \right] = 1.
    \end{align}
\end{corollary}

Next, Theorem \ref{thm:PDP_ratio} implies that DP is equivalent to a prior-independent bound on an adversary's posterior-to-prior ratio holding with probability 1.

\begin{corollary} \label{cor:pure_DP_ratio}
    $M$ satisfies $(\varepsilon,0)$-DP if and only if in the setting described in Section \ref{sec:setting}, for any database $\data$, target $i$, $p_i \in (0,1]$, and under both $Y \leftarrow M(\data)$ and $Y \leftarrow M(\data_{-i})$, the distribution of $R_i$, as defined in Theorem \ref{thm:PDP_ratio}, satisfies
    \begin{align} \label{eq:pure_DP_ratio}
        P\left[e^{-\varepsilon} \leq R_i \leq e^{\varepsilon} \right] = 1.
    \end{align}
\end{corollary}

Finally, Theorem \ref{thm:PDP_diff} implies that DP is equivalent to a prior-independent bound on an adversary's posterior-to-prior difference holding with probability 1.

\begin{corollary} \label{cor:pure_DP_diff}
    $M$ satisfies $(\varepsilon,0)$-DP if and only if in the setting described in Section \ref{sec:setting}, for any database $\data$, target $i$, $p_i \in [0,1]$, and under both $Y \leftarrow M(\data)$ and $Y \leftarrow M(\data_{-i})$, the distribution of $D_i$, as defined in Theorem \ref{thm:PDP_diff}, satisfies
    \begin{align} \label{eq:pure_DP_diff}
        P\left[-\frac{e^{\varepsilon/2} - 1}{e^{\varepsilon/2} + 1} \leq D_i \leq \frac{e^{\varepsilon/2} - 1}{e^{\varepsilon/2} + 1} \right] = 1.
    \end{align}
\end{corollary}

\subsection{Results for Approximate DP} \label{sec:results_approx_DP}

We now examine the corollaries of our results involving approximate DP. 
Although we state the results as corollaries of the results in Section \ref{sec:results_PDP}, these will be the primary results we leverage for the applications in Section \ref{sec:applications}.

To apply the results from Section \ref{sec:results_PDP} relating PDP to quantities involving disclosure risk, we must first relate PDP to approximate DP. Theorem \ref{thm:PDP_to_DP} demonstrates that mechanisms that satisfy $(\varepsilon, \delta)$-PDP also satisfy $(\varepsilon, \delta)$-DP. Theorem \ref{thm:DP_to_PDP} presents a partial converse: mechanisms that satisfy $(\varepsilon, \delta)$-DP satisfy $(\varepsilon', \delta')$-PDP for a curve of $\delta' > \delta$ and $\varepsilon' > \varepsilon$.

\begin{theorem} \label{thm:DP_to_PDP}
    If $M$ satisfies $(\varepsilon, \delta)$-DP, then for any $\delta' \in (\delta, 1]$, $M$ satisfies $(\varepsilon', \delta')$-PDP for $\varepsilon' = \log(\delta'e^{\varepsilon} +\delta) - \log(\delta' - \delta)$.
\end{theorem}

%A related result appears in \cite{zhao2019reviewing}, although this work is not peer reviewed. We present a complete, novel proof of Theorem \ref{thm:DP_to_PDP} in Appendix \ref{proof:DP_to_PDP}. 
See Appendix \ref{proof:DP_to_PDP} for a full proof of Theorem \ref{thm:DP_to_PDP}. We acknowledge that a similar result appears in \cite{zhao2019reviewing}; in an effort to be self-contained, we present a proof that does not utilize the result in \cite{zhao2019reviewing}.

\begin{figure*}[t]
    \centering
    \includegraphics[width=2.1\columnwidth]
    {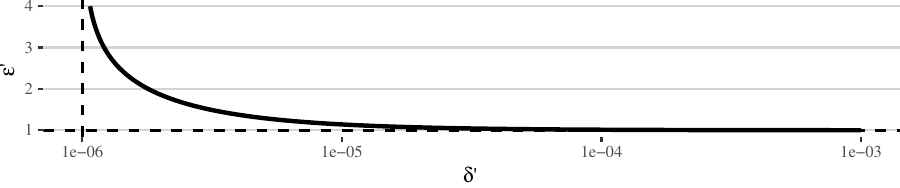}
    \caption{If $M$ satisfies $(\varepsilon = 1, \delta = 10^{-6})$-DP, then $M$ satisfies $(\varepsilon', \delta')$-PDP for points along the above solid curve. $\varepsilon$ and $\delta$ are represented by the dashed lines.}
    \label{fig:DP_to_PDP}
    \Description{When $\delta'$ (on the x-axis) is between $\delta = 10^{-6}$ and $10^{-5}$, $\varepsilon'$ (on the y-axis) is much larger than $\varepsilon = 1$. But as $\delta'$ increases beyond this range, $\varepsilon' \to \varepsilon = 1$.}
\end{figure*}

Notably, when $\delta' \gg \delta$, it follows that $\varepsilon' \approx \varepsilon$. This is demonstrated in Figure \ref{fig:DP_to_PDP} for a mechanism satisfying ($\varepsilon = 1, \delta = 10^{-6})$-DP. It is generally recommended that $\delta$ be cryptographically small, but it may be reasonable for $\delta'$---which represents a failure rate for a bound on an adversary's posterior---to be larger. For example, we might set $\delta'$ to be $0.05$ or $0.01$, as is typical for false positive rates in applied statistics. For cryptographically small $\delta$, this would imply that $\varepsilon' \approx \varepsilon$.

Next, we demonstrate the relationship between approximate DP and a prior-dependent bound on an adversary's posterior probability. Proofs of this and all subsequent results in this section are presented in Appendix \ref{proof:approx_DP}.

\begin{corollary} \label{cor:approx_DP_posterior}
    In the setting described in Section \ref{sec:setting},
    \begin{enumerate}
        \item If $M$ is such that for any database $\data$, target $i$, $p_i \in [0,1]$, and under both $Y \leftarrow M(\data)$ and $Y \leftarrow M(\data_{-i})$, the distribution of $X_i$ satisfies
        \begin{align} \label{eq:post_to_eps_delta}
            P\left[\frac{p_i}{p_i + (1-p_i)e^{\varepsilon}} \leq X_i \leq \frac{p_i}{p_i + (1-p_i)e^{-\varepsilon}} \right] \geq 1-\delta,
        \end{align}
        then $M$ satisfies $(\varepsilon, \delta)$-DP.

        \item If $M$ satisfies $(\varepsilon, \delta)$-DP, then for any $\delta' > \delta$, database $\data$, target $i$, $p_i \in [0,1]$, and under both $Y \leftarrow M(\data)$ and $Y \leftarrow M(\data_{-i})$, for $\varepsilon' = \log(\delta'e^{\varepsilon} +\delta) - \log(\delta' - \delta)$, the distribution of $X_i$ satisfies
        \begin{align} \label{eq:eps_delta_to_post}
            P\left[\frac{p_i}{p_i + (1-p_i)e^{\varepsilon'}} \leq X_i \leq \frac{p_i}{p_i + (1-p_i)e^{-\varepsilon'}} \right] \geq 1-\delta'.
        \end{align}
    \end{enumerate}
\end{corollary}

We now examine the relationship between the posterior-to-prior ratio and approximate DP.

\begin{corollary} \label{cor:approx_DP_ratio}
    In the setting described in Section \ref{sec:setting}, let $R_i$ be as defined in Theorem \ref{thm:PDP_ratio}.
    \begin{enumerate}
        \item If $M$ is such that for any database $\data$, target $i$, $p_i \in (0,1]$, and under both $Y \leftarrow M(\data)$ and $Y \leftarrow M(\data_{-i})$, $R_i$ satisfies
        \begin{align} \label{eq:ratio_to_eps_delta}
            P\left[e^{-\varepsilon} \leq R_i \leq e^{\varepsilon} \right] \geq 1-\delta,
        \end{align}
        then $M$ satisfies $(\varepsilon, \delta)$-DP.

        \item If $M$ satisfies $(\varepsilon, \delta)$-DP, then for any $\delta' > \delta$, database $\data$, target $i$, $p_i \in (0,1]$, and under both $Y \leftarrow M(\data)$ and $Y \leftarrow M(\data_{-i})$, for $\varepsilon' = \log(\delta'e^{\varepsilon} +\delta) - \log(\delta' - \delta)$,
        \begin{align} \label{eq:eps_delta_to_ratio}
            P\left[e^{-\varepsilon'} \leq R_i \leq e^{\varepsilon'} \right] \geq 1-\delta'.
        \end{align}
    \end{enumerate}
\end{corollary}

Finally, we examine the relationship between the posterior-to-prior difference and approximate DP.

\begin{corollary} \label{cor:approx_DP_diff}
    In the setting described in Section \ref{sec:setting}, let $D_i$ be as defined in Theorem \ref{thm:PDP_diff}.
    \begin{enumerate}
        \item If $M$ is such that for any database $\data$, target $i$, $p_i \in [0,1]$, and under both $Y \leftarrow M(\data)$ and $Y \leftarrow M(\data_{-i})$, the distribution of $D_i$ satisfies
        \begin{align} \label{eq:diff_to_eps_delta}
            P\left[-\frac{e^{\varepsilon/2} - 1}{e^{\varepsilon/2} + 1} \leq D_i \leq \frac{e^{\varepsilon/2} - 1}{e^{\varepsilon/2} + 1} \right] \geq 1-\delta,
        \end{align}
        then $M$ satisfies both $(\varepsilon, 2\delta)$-DP and $(\tilde{\varepsilon}, \delta)$-DP for
        \begin{align*}
            \tilde{\varepsilon} = \log(3e^{\varepsilon/2} - 1) - \log(3 - e^{\varepsilon/2}).
        \end{align*}

        \item If $M$ satisfies $(\varepsilon, \delta)$-DP, then for any $\delta' > \delta$, database $\data$, target $i$, $p_i \in [0,1]$, and under both $Y \leftarrow M(\data)$ and $Y \leftarrow M(\data_{-i})$, for $\varepsilon' = \log(\delta'e^{\varepsilon} +\delta) - \log(\delta' - \delta)$,
        \begin{align} \label{eq:eps_delta_to_diff}
            P\left[-\frac{e^{\varepsilon'/2} - 1}{e^{\varepsilon'/2} + 1} \leq D_i \leq \frac{e^{\varepsilon'/2} - 1}{e^{\varepsilon'/2} + 1} \right] \geq 1-\delta'.
        \end{align}
    \end{enumerate}
\end{corollary}

\section{Applications} \label{sec:applications}

In this section, we demonstrate the versatility of the results from Section \ref{sec:results}. First, we construct explanations about aspects of DP via these results for audiences with varying technical backgrounds. Second, we interpret DP composition theorems through the lens of these results. Third, we generalize the work of \cite{kazan2024prior} on selecting $\varepsilon$ for DP. Finally, we examine use of these results to determine the worst-case prior probability from the perspective of a data curator.

\subsection{Understanding DP} \label{sec:understand_DP}

Prior work (e.g., \cite{cummings2021need, franzen2022private, nanayakkara2023chances}) suggests that explanations based on risk can be effective for conveying DP guarantees (see \cite{dibia2024sok} for a review of the literature on communicating privacy guarantees). In this section, we demonstrate how the results of Section \ref{sec:results} can be used to craft novel explanations, leveraging findings in the literature regarding what explanations are most successful. In particular, \cite{franzen2022private} finds that the audience's technical background is an important factor in the effectiveness of an explanation. To mitigate this issue, we consider separate explanations for two audiences. The first is a nontechnical audience, e.g., a stakeholder in a project involving DP or a participant in a DP data release. The second is a technical audience, e.g., 
a data curator performing a DP release or an analyst receiving privatized data. We assume the technical audience has working knowledge of elementary statistics, particularly the basics of Bayesian analysis.

First, we consider direct explanation of the approximate DP guarantee. For a nontechnical audience, we aim for the explanation to be intuitive but not necessarily technically precise, as follows. 

\begin{zquote}
    If I contribute my data to a dataset analyzed via differential privacy, then after a knowledgeable adversary sees the result, with high probability, their beliefs about my presence in the dataset can change by only a set, pre-determined amount.
\end{zquote}

We now present the explanation for a technical audience, which we aim to be technically precise but avoid the inclusion of potentially confusing formulas. 
\begin{zquote}
    Suppose an adversary believes there is probability $p_i$ I contributed to a particular dataset. Consider a worst-case scenario: the adversary possesses everyone in the data's complete information; all they do not know is whether I am present. If a release based on this dataset satisfies ($\varepsilon,\delta$)-DP, then with high probability, after the adversary sees the release, $p_i$ can change by only a pre-determined amount. Larger values of $\varepsilon$ and $\delta$ allow for larger changes in $p_i$.
\end{zquote}

Audiences with sufficient technical background can be referred to (\ref{eq:eps_delta_to_post}) and Corollary \ref{cor:approx_DP_posterior}, which provide the bounds on an adversary's change in beliefs and the formal theorem statement, respectively.

Other concepts in DP can also be explained intuitively via our results. For example, consider explanation of the difference between pure and approximate DP. Understanding the differences between these privacy guarantees might be very important to, for example, regulators evaluating the privacy protections of DP systems. But a regulator may not have sufficient technical background to understand the formal definitions in Section \ref{sec:background}. The results in Section \ref{sec:results} provide a clean way of distinguishing the guarantees.\footnote{A similar explanation can be used to distinguish pure DP and zCDP via the conversion of zCDP to approximate DP in Theorem \ref{thm:zCDP_to_approx_DP}.} For the nontechnical audience, we provide the following explanation, which builds upon the previous. 

\begin{zquote}
    Under approximate DP, it is unlikely, but not impossible, that the adversary's beliefs change by a large amount after seeing the result. Under pure DP, this cannot occur; it is guaranteed that their beliefs can change by only a set, pre-determined amount.
\end{zquote}

For the technical audience, we provide the following explanation, again building on the previous.

\begin{zquote}
    Under approximate $(\varepsilon, \delta)$-DP, with high probability, the amount the adversary's prior probability, $p_i$, can change is bounded by a quantity depending on $\varepsilon$ and $\delta$. Potentially, the change in $p_i$ can exceed this bound. Under pure $\varepsilon$-DP, however, it is guaranteed that the change cannot exceed the bound.
\end{zquote}

Audiences with more technical background can be referred to Corollaries \ref{cor:pure_DP_posterior} and \ref{cor:approx_DP_posterior}, which provide the bounds on the change in an adversary's beliefs.

Finally, we return to the example from Section \ref{sec:intro}. The example is reproduced below.

\begin{repexample}{ex:simple}
    Employees are required to evaluate their manager, including answering the \texttt{YES}/\texttt{NO} question \textit{``Do you feel adequately supported by your manager?''} Managers will receive a summary with the count of each response. There is concern that if an employee believes their manager knows all others will respond \texttt{YES}, they risk retaliation should they respond \texttt{NO}. To avoid this scenario, noise is added to summaries before they are released to the managers. The employees and managers are not provided with the details of the noise mechanism, but are told it satisfies ($\varepsilon = 0.1, \delta = 10^{-7})$-DP.
\end{repexample}

Leveraging the results in Section \ref{sec:results},\footnote{Note that although Example \ref{ex:simple} is stated as an attribute inference attack, we can interpret the worst-case as a membership inference attack, as follows. We let $x$ represent the database of employees that respond \texttt{NO} to the survey. In the worst-case for the employee, the manager knows that all other employees responded \texttt{YES}. Thus, in the worst-case,the employee's response corresponds to membership in $x$ and we may leverage the results of Section \ref{sec:results}.} if the manager believes there is a 50\% probability a particular employee will respond \texttt{NO} a priori, then in the case where $\delta = 0$ (by Corollary \ref{cor:pure_DP_posterior}) the probability the manager assigns to the employee responding \texttt{NO} is guaranteed to be between 48\% and 52\% after observing the summary counts. In Example \ref{ex:simple}, however, $\delta = 10^{-7}$, so we must leverage Corollary \ref{cor:approx_DP_posterior}. If we take $\delta' = 0.01$, then since $\delta' \gg \delta$, it follows that $\varepsilon' \approx \varepsilon$. Thus, with high (99\%) probability, the probability the manager assigns to the employee responding \texttt{NO} is guaranteed to be between 48\% and 52\% after observing the summary counts.

%As discussed in Section \ref{sec:intro}, if the manager believes there is a 50\% probability a particular employee will respond \texttt{NO} before viewing the summary counts and the manager were to know all other employees responded \texttt{YES}, then with 99\% probability, the probability the manager assigns to the employee responding \texttt{NO} will be between 48\% and 52\% after observing the summary counts. These bounds are computed via (\ref{eq:eps_delta_to_post}). In contrast, if the manager knows all other employees responded \texttt{YES}, then without the addition of DP noise the manager's probability the employee responded \texttt{NO} would be guaranteed to be either 0\% or 100\%.

More generally, we can avoid making assumptions about the manager's initial probability the employee will respond \texttt{NO}. By Corollary \ref{cor:approx_DP_ratio}, with 99\% probability, the manager's initial probability can increase by at most a factor of 1.1 or decrease by at most a factor of 0.90. Similarly, by Corollary \ref{cor:approx_DP_diff}, with 99\% probability, the manager's initial probability can increase or decrease by at most 2\% in absolute terms. Both of these bounds hold no matter what the manager's probability was initially. If the company wishes to describe the privacy guarantee when asking employees to provide their responses, then they might---adhering to the guidance of prior work---summarize the above as follows.

\begin{zquote}
    Aggregate results from this survey will be released to your manager under differential privacy. Thus, even if you are the only employee to reply \texttt{NO}, your manager will not be certain you did so. It is unlikely the probability your manager assigns to you responding \texttt{NO} will change by more than 2\%.
\end{zquote}

\subsection{Examining DP Composition}

\begin{figure*}[t]
    \centering
    \begin{minipage}[t]{\columnwidth}
        \centering
        \includegraphics[width=\columnwidth]{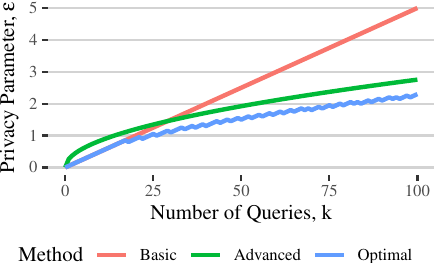}
        \caption{The total privacy parameter, $\varepsilon$, as a function of $k$ for the composition of $k$ pure DP mechanisms, each with $\varepsilon_j = 0.05$. We take $\delta = 10^{-6}$. Colors correspond to composition methods.}
        \label{fig:DP_comp_eps}
        \Description{Plot of the privacy parameter, $\varepsilon$, as a function of the number of queries, $k$, for basic, advanced, and optimal composition. The $\varepsilon$ for basic composition increases linearly, while the $\varepsilon$ for the other two increase sub-linearly.}
    \end{minipage}%
    \hspace{7.5mm}
    \begin{minipage}[t]{\columnwidth}
        \centering
        \includegraphics[width=\columnwidth]{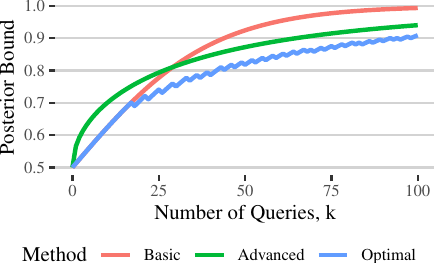}
        \caption{Posterior upper bound as a function of $k$ for the composition of $k$ $(\varepsilon_j = 0.05)$-DP mechanisms. We take $\delta = 10^{-6}$ and $p_i = 0.5$. Colors correspond to composition methods.}
        \label{fig:DP_comp_post}
        \Description{Plot of the posterior bound as a function of the number of queries, $k$, for basic, advanced, and optimal composition. The posterior bound for basic composition increases quickly and tapers off to one, while the posterior bound for the other two increase more slowly.}
    \end{minipage}
\end{figure*}

The semantic characterizations of approximate DP also provide an interesting, alternate perspective on DP composition results. Often, composition results are visualized as in Figure \ref{fig:DP_comp_eps}, which presents the composition of $k$ pure DP mechanisms, each with $\varepsilon_j = 0.05$. The $(\varepsilon, \delta = 10^{-6})$ guarantee of three standard DP composition methods is presented as a function of $k$. However, audiences without background in formal privacy may not find this presentation intuitive. It is also unclear how many queries are required to raise concerns about individuals' disclosure risk.

%In this section, we examine the DP composition results discussed in Section \ref{sec:DP} from two perspectives. First, continuing Section \ref{sec:understand_DP}'s theme of interpretability, we present a novel adversarial interpretation of DP composition. Second, we examine bounds on an adversary's posterior for a realistic series of zCDP mechanisms.

%To begin, consider the composition of $k$ pure DP mechanisms, each with $\varepsilon_j = 0.05$. Figure \ref{fig:DP_comp_eps} presents a standard comparison of the $(\varepsilon, \delta = 10^{-3})$ guarantee from the three standard DP composition methods reviewed in Theorem \ref{thm:DP_comp} as a function of the number of queries, $k$. We see that the $\varepsilon$ from basic composition increases linearly in $k$, while the $\varepsilon$ from advanced and optimal composition increase sub-linearly. This analysis, however, may be difficult for audiences without background in formal privacy to understand. It is also unclear after how many queries concerns should arise regarding individuals' disclosure risk.

We reconsider this scenario from the adversarial perspective of Section \ref{sec:results}. Suppose all releases are from a common database and the adversary has 50\% prior probability of a target being in this database. Figure \ref{fig:DP_comp_post} examines the upper bound on the adversary's posterior probability as a function of $k$, computed via Corollary \ref{cor:approx_DP_posterior} with $\delta' = 0.05$ and the $\varepsilon$ from Theorem \ref{thm:DP_comp}. We see, for example, that with 95\% probability, the $k$ required for the upper bound on the adversary's posterior probability to reach 80\% differs substantially between the different composition methods. The 80\% threshold is exceeded after $28$ queries for basic, $51$ queries for advanced, and $96$ queries for optimal composition. This provides a meaningful method for determining the number of queries required to exceed a desired disclosure risk bound. Notably, we are aware of no prior work able to provide posterior semantics with the advanced or optimal composition theorems; the bound from basic composition is the best that can be achieved from, e.g., \cite{lee2011much, kazan2024prior, kifer2022bayesian}.

We now examine the composition properties of the type of releases currently performed by many organizations.%\footnote{We aim for the setting and privacy parameters in this example to be generally representative of current industry practices. This example is not based on the methods used by any particular organization.}

\begin{example} \label{ex:zCDP}
    An organization uses zCDP to release daily statistics about individuals utilizing a particular service. They report a privacy budget of $\rho = 0.01$ with a privacy unit of user-day. The service requires an annual subscription, so the underlying database is roughly constant throughout the year. 
    %Since many individuals are daily users of the service, 
    There is concern that after many daily releases, an adversary could become confident that a target is or is not a user of the service.
\end{example}

\begin{figure*}[t]
    \centering
    \begin{minipage}[t]{\columnwidth}
        \includegraphics[width=\columnwidth]{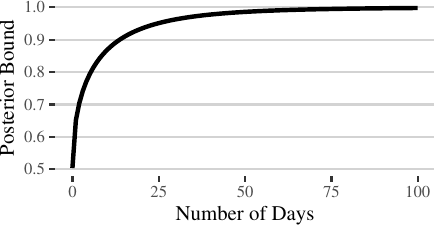}
        \caption{The posterior upper bound of an adversary with prior probability $p_i = 0.5$ as a function of the number of releases. We set $\delta' = 0.01$ and each releases satisfies $(\rho = 0.01)$-zCDP.}
        \label{fig:zCDP_comp_post}
        \Description{The maximum posterior increases quickly from 0.5 as the number of days increases from 0 to 50. It converges to one as the number of days increases past 50.}
    \end{minipage}%
    \hspace{7.5mm}
    \begin{minipage}[t]{\columnwidth}
        \includegraphics[width=\columnwidth]{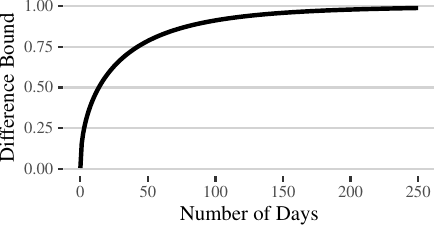}
        \caption{The maximum posterior-to-prior difference over all prior probabilities $p_i$ as a function of the number of releases. We set $\delta' = 0.01$ and each releases satisfies $(\rho = 0.01)$-zCDP.}
        \label{fig:zCDP_comp_diff}
        \Description{The maximum difference increases quickly from 0 as the number of days increases from 0 to 150. It converges to one as the number of days increases past 150.}
    \end{minipage}
\end{figure*}

Suppose an adversary has a 50\% prior probability the target uses the service and complete knowledge of all other individuals that use the service. Figure \ref{fig:zCDP_comp_post} plots a 99\% upper bound on the adversary's posterior probability as a function of the number of days elapsed. Even with a small daily $\rho$, the bound increases quickly. It reaches 83\% after one week, 96\% after one month, and exceeds 99\% after 58 days. Of course, the choice of an adversary with $p_i = 0.5$ is arbitrary. We can  avoid making assumptions about this quantity by instead examining bounds on an adversary's posterior-to-prior difference across all potential $p_i$. Figure \ref{fig:zCDP_comp_diff} plots the 99\% upper bound on the posterior-to-prior difference as a function of days elapsed. The bound on the posterior-to-prior difference is 38\% after one week, 67\% after one month, and exceeds 98\% after 202 days. This corresponds to worst-case increases of 31\% to 69\%, 19\% to 81\%, and 1\% to 99\%, respectively.

We emphasize that the quantities in Figures \ref{fig:DP_comp_post}-\ref{fig:zCDP_comp_diff} are upper bounds. Throughout, we assume a worst-case, likely unrealistically knowledgeable adversary. Additionally, we examine the largest possible increase from prior to posterior probability under \textit{any} $(\varepsilon, \delta)$-DP mechanism. It can be shown that specific mechanisms---for example, the Laplace mechanism under pure DP---cannot achieve these upper bounds. If the form of the privacy loss random variable is known for a particular mechanism, Theorem \ref{thm:PLRV_posterior} can be used to compute tighter bounds on the adversary's posterior distribution. We also note that many mechanisms, such as the Gaussian mechanism, produce a curve of $(\varepsilon, \delta)$-DP guarantees. In such cases, different points along the curve will yield different disclosure risk interpretations.
Finally, we note that for Example \ref{ex:zCDP}, analysts may be able to leverage recent results, e.g., \cite[Sec.~7.4]{kifer2022bayesian} or \cite[App.~F]{zhu2022optimal} to derive tighter semantics for zCDP. We leave development of these ideas to future work.

\subsection{Setting Privacy Parameters} \label{sec:choosing_eps}

In recent work, Kazan \& Reiter (2024) \cite{kazan2024prior} propose a framework for formalizing the process of setting $\varepsilon$ for a release under pure DP, making the independence assumption discussed in Section \ref{sec:prior_work}. The authors suggest organizations produce disclosure risk profiles, which represent the maximum posterior-to-prior ratio they are willing to incur from adversaries with various initial prior probabilities. The authors demonstrate this profile uniquely implies a maximum allowable $\varepsilon$ for the release. Notably, since \cite{kazan2024prior} considers only pure DP, these results can be used only with basic composition. The results are also agnostic to the particular DP mechanism used for the release. 

The results in Section \ref{sec:results} can be used to construct similar frameworks for approximate DP in the setting described in Section \ref{sec:setting}. For example, suppose an organization has a desired $\delta$ (e.g., using the recommendation of \cite{kasiviswanathan2014semantics}), but is uncertain what $\varepsilon$ to use for a release. They can produce a disclosure risk profile, as described in \cite{kazan2024prior}, where they require a posterior-to-prior ratio bound holds with $1-\delta'$ probability, for some desired $\delta'$. The allowable $\varepsilon$ can then be computed analogously to the proposal of \cite{kazan2024prior}. In sequential settings, this framework is advantageous because it can be used to set $\varepsilon$ via stronger composition theorems. For mechanisms with closed form privacy loss random variables, Theorem \ref{thm:PLRV_posterior} can be used to examine the adversary's exact distribution of posterior probabilities, rather than using the upper bound over all DP mechanisms from Theorem \ref{cor:approx_DP_posterior}. The following example illustrates the use of this framework.

%We consider an example to illustrate the use of this framework.

\begin{example}
    A government agency plans to disclose monthly summary statistics regarding queries of a secure government database. They will use DP for each release with a common $\varepsilon$ and $\delta = 10^{-8}$ with user-month as the privacy unit. Access to the database is restricted, with only a few individuals having approval. Over the course of a year, they wish that, with at least 99\% probability, the probability an adversary assigns to any particular person being on the approved list will not increase by more than 20\%. They desire to have a total $\delta = 10^{-6}$.
\end{example}

The requirement that an adversary's posterior-to-prior difference not exceed 20\% implies a disclosure risk profile. Straightforwardly\footnote{From part 2 of Corollary \ref{cor:approx_DP_diff}, $(e^{\varepsilon'/2} - 1)/(e^{\varepsilon'/2} + 1) = 0.2$ implies $\varepsilon' = 0.81$. Solving $0.81 = \log(\delta'e^{\varepsilon} +\delta) - \log(\delta' - \delta)$ yields $\varepsilon = 0.81$.}, this risk profile implies a total privacy budget of $\varepsilon_{\textrm{total}} \geq 0.81$.
%  By Corollary \ref{cor:diff_eps}, the bound $\tilde{d} = 0.2$ implies a total privacy budget of
% \begin{align*}
%     \varepsilon_{\textrm{total}} \leq 2\log(1 + \tilde{d}) - 2\log(1-\tilde{d}) \approx 0.811.
% \end{align*}
We perform a simple optimization to determine the maximal $\varepsilon$ such that the overall guarantee is $(\varepsilon_{\textrm{total}}, \delta = 10^{-6})$-DP and each release satisfies $(\varepsilon, \delta = 10^{-8})$-DP via the optimal composition theorem (Theorem \ref{thm:DP_comp}). This yields $\varepsilon \approx 0.135$ for each release.

Notably, the framework of Kazan \& Reiter (2024) \cite{kazan2024prior} is unable to incorporate $\delta > 0$ into the recommendation. For comparison, if we take $\delta = 0$ throughout and perform the same analysis via the original framework with pure DP and basic composition, the recommendation is $\varepsilon \approx 0.068$ per release. This represents a factor of two increase in the $\varepsilon$ recommendation due to our generalization of this framework. We emphasize, however, that this is not an apples-to-apples comparison due to differences in assumptions.

\subsection{Worst-case Adversary Prior Probabilities} \label{sec:worst_case}

In Section \ref{sec:results}, we derived prior-independent bounds for the posterior-to-prior ratio and difference by finding the $p_i$ that yield the largest changes. %These $p_i$ can be useful in practice for settings where analysis of a DP mechanism's disclosure risk requires making assumptions about an adversary's prior probability. 
However, for some analyses it can be useful to select a particular value of $p_i$. A natural choice for this value is the $p_i$ yielding the largest change in a desired criteria.
Consider the following demonstrative example.

%Analyses involving Bayesian semantics of DP may require making assumptions about an adversary's prior probability. As a demonstrative example, consider the following setting.

\begin{example} \label{ex:worst_case}
    A technology company's services are known to be used by 10\% of the population. The company possesses a large database containing information about all their users and opts to implement a complex DP system based on DP stochastic gradient descent \cite{abadi2016deep} to protect users' privacy when analysts query the database. They report a total $\varepsilon = 1.8$ and $\delta = 10^{-5}$ over a trial period. The company's data governance team requests a worst-case disclosure risk analysis based on the current implementation.
\end{example}

Suppose analysts wish to assess the potential increase in an adversary's beliefs about whether a target is a user of the company's services, focusing on potential targets in the database. As demonstrated by McClure \& Reiter (2012) \cite{mcclure2012differential}, the reported disclosure risk will be sensitive to the prior probability used for analysis. We consider a few selections for this quantity. One option is to set the adversary's prior probability to 50\% (as in, e.g., \cite{kazan2025assessing, nanayakkara2023chances}), implying they believe it is equally likely the target is and is not a user. Then with 95\% probability, by the results in Section \ref{sec:results_approx_DP}, the adversary's posterior probability will be at most 86\%, representing a difference of 36\% and a ratio of $1.7$. Another option is to set the adversary's prior probability to 10\%, implying they know the target is in the population, but little else. This implies that the adversary's posterior probability will be at most 40\%, representing a difference of 30\% and a ratio of $4.0$.

To supplement this analysis, analysts may wish to consider a notion of the worst-case $p_i$. Of interest are the $p_i$ that maximize disclosure risk criteria, e.g., the posterior-to-prior ratio and difference. However, recall that $p_i$ and $X_i$ represent an adversary's probability that a target is present in a database ($I_i = 1$). %As suggested by \cite{lee2011much},
In some settings, a data curator may be most concerned about the probability that a target is not present in a database ($I_i = 0$); see \cite{lee2011much}. This corresponds to bounding the change from $1-p_i$ to $1-X_i$. To account for all settings, we examine the maximum of these two changes.

\begin{figure*}
    \begin{minipage}[t]{\columnwidth}
        \includegraphics[width=\columnwidth]{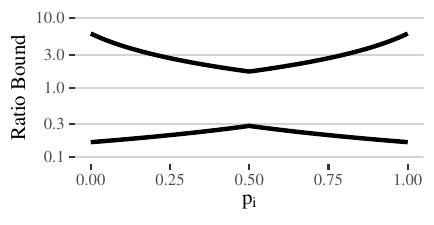}
        \caption{Worst-case bounds an adversary's posterior-to-prior ratio for $(\varepsilon = 1.8, \delta=10^{-5})$-DP as a function of $p_i$. The bound holds with at least 95\% probability. Note the log-scale on the y-axis.}
        \label{fig:ratio_plot}
        \Description{The ratio bounds are largest when $p_i$ is near $0$ and $1$, reaching a range of nearly $[0.1, 10]$. The bounds are smallest when $p_i = 0.5$, approximately $[0.3, 3]$.}
    \end{minipage}%
    \hspace{7.5mm}
    \begin{minipage}[t]{\columnwidth}
        \includegraphics[width=\columnwidth]{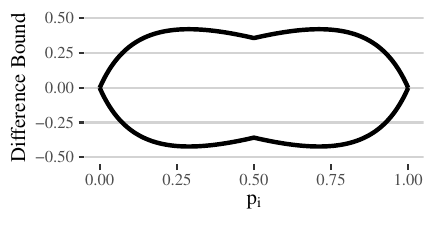}
        \caption{Worst-case bounds an adversary's posterior-to-prior difference for $(\varepsilon = 1.8, \delta = 10^{-5})$-DP as a function of $p_i$. The bound holds with at least 95\% probability.}
        \label{fig:diff_plot}
        \Description{The difference bounds are largest when $p_i$ is close to, but not equal to 0.5, reaching a range of nearly $[-0.5, 0.5]$. The bounds collapse to 0 when $p_i$ is near $0$ and $1$.}
    \end{minipage}
\end{figure*}

We first examine the maximal posterior-to-prior ratio the adversary can achieve. By Corollary \ref{cor:approx_DP_posterior}, the bounds for $I_i = 1$ are $1/(p_i + (1-p_i)e^{\varepsilon'}) \leq X_i/p_i \leq 1/(p_i + (1-p_i)e^{-\varepsilon'})$, where $\varepsilon' = \log(\delta'e^{\varepsilon} +\delta) - \log(\delta' - \delta)$. We can similarly bound the analogous quantity for $I_i = 0$, which is $1/(p_ie^{\varepsilon'} + (1-p_i)) \leq (1-X_i)/(1-p_i) \leq 1/(p_ie^{-\varepsilon'} + (1-p_i))$.
These bounds hold with $1-\delta'$ probability. 
Figure \ref{fig:ratio_plot} plots 95\% probability bounds on $\tilde{R}_i = \max\{X_i/p_i, (1-X_i)/(1-p_i)\}$ as a function of $p_i$.
We see that the worst-case adversary by this metric is one with $p_i$ close to $0$ or $1$. Such an adversary can achieve the largest posterior-to-prior ratio, nearing the bound in Corollary \ref{cor:approx_DP_ratio} of $e^{\pm\varepsilon}$. 
This implies adversaries that achieve large posterior-to-prior ratios must begin with an extreme, incorrect prior belief. For example, for an adversary's probability that $I_i = 1$ to increase by a factor close to $e^\varepsilon$, it must be that $p_i \approx 1$. Whether such an adversary achieving a large posterior-to-prior ratio is of practical concern is context dependent. In the setting of Example \ref{ex:worst_case}, an adversary with prior probability near zero can achieve a posterior-to-prior ratio of at most $6$.

We can similarly examine the maximal posterior-to-prior difference the adversary described in Section \ref{sec:setting} can achieve. By Corollary \ref{cor:approx_DP_posterior}, for $I_i = 1$, $p_i/(p_i + (1-p_i)e^{\varepsilon'}) - p_i \leq X_i - p_i \leq p_i/(p_i + (1-p_i)e^{-\varepsilon'}) - p_i$, where $\varepsilon' = \log(\delta'e^{\varepsilon} +\delta) - \log(\delta' - \delta)$. For $I_i = 0$, since $(1-X_i) - (1-p_i) = -(X_i - p_i)$, bounds are obtained by negating the above. Let $\tilde{D}_i = \max\{X_i - p_i, (1-X_i) - (1-p_i)\}$ and again consider a release satisfying $(\varepsilon = 2, \delta = 10^{-6})$-DP for $\delta' = 0.01$. Figure \ref{fig:diff_plot} plots the bounds on $\tilde{D}_i$ as a function of $p_i$, which hold with 99\% probability. %For $p_i < 0.5$, the adversary's potential increase in the probability $I_i = 1$ is larger than that of $I_i = 0$. The adversary's potential decrease in the probability $I_i = 0$, however, is larger than that of $I_i = 1$. For $p_i > 0.5$, the reverse is true. 
We see that the worst case adversaries are those with $p_i = 0.27$ and $p_i = 0.73$, having a maximum change from prior to posterior of $\pm 0.46$. This corresponds to the result of Lemma \ref{lem:diff_worst}, which indicates the bounds are most extreme when $p_i = 1/(1+e^{\pm\varepsilon'/2})$ and the worst-case bound is $\pm (e^{\varepsilon'/2}-1)/(e^{\varepsilon'/2}+1)$. As $\varepsilon \to 0$, the worst-case $p_i \to 0.5$. This provides a novel $p_i$ to focus an analysis. By Lemma \ref{lem:diff_worst}, adversaries with any other $p_i$ are unable to achieve as large a change from prior to posterior probability as an adversary with $p_i = 1/(1+e^{\pm\varepsilon'/2})$. Focusing on these $p_i$ seems in line with the recommendation of \cite{hotz2022balancing}, who argue that individuals in a sensitive dataset should be most concerned about controlling an adversary's posterior-to-prior difference. In the setting of Example \ref{ex:worst_case}, an adversary with $p_i = 1/(1+e^{\varepsilon'/2}) \approx 29\%$ can achieve a posterior probability at most 71\%, representing a difference of 42\% and a ratio of $2.5$.

Generally, whether incorporating a particular $p_i$ makes sense is context dependent. For example, an adversary with $p_i = 0.5$ seems reasonable for Example \ref{ex:worst_case}. If, however, a target's presence in a database implies they have a rare disease, then an adversary having $p_i = 0.5$ would mean they believe the target is much more likely to have the disease than a person randomly sampled from the population. I.e., this may represent an adversary using auxiliary information to inform their initial beliefs. In the interest of transparency, it may be advisable for analysts to consider adversaries with a variety of prior probabilities. The worst-case prior probabilities provide values that can be considered in any setting.

\section{Discussion} \label{sec:discuss}

We now summarize this work, provide concluding thoughts, and discuss the work's limitations and future directions.

\subsection{Overview}

% To begin, we summarize our work and suggest some of its potential implications.

% \subsubsection{Summary}

In this work, we present a series of Bayesian sematic characterizations of probabilistic, pure, and approximate DP. We demonstrate that PDP is equivalent to both bounding an adversary's posterior probability and to bounding an adversary's posterior-to-prior ratio. We also demonstrate that pure DP is equivalent to bounding an adversary's posterior-to-prior difference. Of practical interest, we characterize approximate DP in terms of bounds on an adversary's posterior probability, posterior-to-prior ratio, and posterior-to-prior difference. As far as we are aware, all of these results are novel. We provide in-depth examples demonstrating potential applications of these results, including explanation of DP to audiences with varying technical background, interpretation of DP composition, selection of the privacy budget for approximate DP, and examination of what a data curator might consider a worst-case adversary's prior probability.

%In this work, we present a Bayesian semantic characterization of $(\varepsilon,\delta)$-differential privacy. Our main results relate privacy loss random variables to an adversary's posterior probability and demonstrate an equivalence between approximate DP and bounds on the distribution of an adversary's posterior probability. We show these results imply bounds an adversary's posterior-to-prior ratio and difference under DP. Prior work generally focused on Bayesian semantics of pure DP and as far as we are aware, this work is the first to provide a characterization of this type for approximate DP. We also provide in-depth examples demonstrating practical applications of these results, including explanation of DP to audiences with varying technical background, interpretation of DP composition theorems, selection of the privacy budget for approximate DP, and examination of what a data curator might consider a worst-case adversary's prior probability.

%\subsubsection{Implications}

These characterizations of DP showcase the strength of the DP definition. To achieve equivalence between pure DP and the three disclosure risk quantities, one must not only assume that the adversary possesses complete knowledge about all individuals in the database except the target, but also that they know the target's attributes. This must then hold for all potential databases and all potential targets in these databases in order for the bound to be equivalent to DP. The fact that such strong assumptions are required demonstrates the powerful privacy protection DP provides. 

%Throughout this work, we suggest a range of practical settings where these results may be useful. This includes an organization understanding the effects of performing sequential data releases, a government agency selecting a privacy budget, and a technology company assessing their potential disclosure risk. These examples demonstrate the versatility of our results in practice.

\subsection{Disclosure Risk Metrics}

It is useful to take a step back and discuss why results regarding disclosure risk metrics like bounds on the adversary’s posterior, posterior-to-prior ratio, and posterior-to-prior difference are needed in addition to existing results about metrics like positive predictive accuracy \cite{thudi2024differential} and comparison to a counterfactual posterior \cite{wood2020designing}. One important consideration is that, since there is a deep line of research on the disclosure risk metrics going back to works like Duncan \& Lambert (1986) \cite{duncan1986disclosure}, audiences with background in statistical disclosure control have decades of experience interpreting disclosure risk in terms of an adversary’s change from prior to posterior. We believe that establishing novel links between DP and these well-studied quantities is particularly useful for audiences with this background---such as decision makers within statistical agencies---who may not be familiar with DP or the other metrics. Additionally, our metrics have useful composition properties (see, e.g., \cite[App.~S5]{kazan2025assessing} in the context of zCDP) and are conducive to accessible explanations for nontechnical audiences.

An additional benefit to these disclosure risk metrics is that they are very adaptable to different use cases, with different metrics being most useful in different contexts. For example, bounds on posterior probabilities are likely most intuitive but require making an assumption about the value of $p_i$, which may not be feasible in some settings. Bounds on posterior-to-prior ratios are likely most familiar to audiences with background in statistical disclosure control and do not require making an assumption about $p_i$. However, these may not be ideal in settings where a small $p_i$ would be expected, such as where membership corresponds to the target possessing a rare characteristic (since a large ratio may correspond to a small absolute increase, e.g., from $p_i = 0.0001$ to $X_i = 0.001$). For this reason, Hotz et al.~(2022) \cite{hotz2022balancing} argues that bounds on posterior-to-prior differences may represent a more meaningful measure of risk than ratios, but this metric is not as well-studied and may be less familiar.

\subsection{Future Work}

We conclude by suggesting avenues for future work.
%\subsubsection{Limitations}
Our results are based on the unbounded, central version of differential privacy and on a strong adversary that possess substantial auxiliary information about the database. Future work might consider analogous results under (1) the bounded version of differential privacy, where neighboring datasets differ in one individual's attributes, rather than one individual's addition or removal, (2) the local version of differential privacy, where randomization is applied at the individual level, or (3) alternate assumptions, such as the independence assumption or counterfactual posterior comparison discussed in Section \ref{sec:prior_work}.
%One avenue for future work considers analogous results under the bounded version of differential privacy, where neighboring datasets differ in one individual's attributes, rather than one individual's addition or removal. Similarly, future work might consider analogous results under the local version of differential privacy, where randomization is applied at the individual level. Another avenue for future work considers analogous results under alternate assumptions, such as by replacing the assumption of a strong adversary with the independence assumption or counterfactual posterior comparison discussed in Section \ref{sec:prior_work}.
%\subsubsection{Other Extensions}
%The implications of this framework regarding the strength of the DP definition also suggest avenues for future work. 
Future work might also consider whether alternate Bayesian semantics apply to a weaker adversary under DP, e.g., an adversary who only possesses complete knowledge about some individuals (as in \cite{dwork2006calibrating}) or a notion of an ``average'' adversary. Alternatively, weakening the assumptions on the adversary in this way may yield a less stringent variant of DP with advantageous properties.

Future work may also further develop our proposed applications.
In Section \ref{sec:understand_DP}, we suggest using our results to explain the DP guarantee. Future work might perform a usability study to assess the efficacy of these explanations.
In Section \ref{sec:choosing_eps}, we suggest using our results to extend \cite{kazan2024prior}'s framework for selecting the privacy budget for pure DP, but we do not fully develop this extension or consider other variations, such as formalizing the selection of $\delta$.
%but we do not provide full details nor do we explore how to incorporate the target's attributes into the analysis (in addition to the target's inclusion). 
Full exploration of this generalization and usability studies to examine its efficacy are both interesting avenues for future work. 
Future work might also explore how mechanisms with closed form PLRVs---e.g., the Gaussian mechanism---may allow for more exact bounds on the adversary's distribution of posterior probabilities. Future work might also incorporate this framework into interactive DP systems, explore its use with other composition theorems, or further develop the ideas regarding worst-case prior probabilities.

\begin{acks}
    Zeki Kazan was partially supported by NSF grant SES-2217456 while conducting this work. 
\end{acks}

\bibliographystyle{ACM-Reference-Format}
\bibliography{references}


\begin{thebibliography}{47}

%%% the Web address.

\ifx \showCODEN    \undefined \def \showCODEN     #1{\unskip}     \fi
\ifx \showDOI      \undefined \def \showDOI       #1{#1}\fi
\ifx \showISBNx    \undefined \def \showISBNx     #1{\unskip}     \fi
\ifx \showISBNxiii \undefined \def \showISBNxiii  #1{\unskip}     \fi
\ifx \showISSN     \undefined \def \showISSN      #1{\unskip}     \fi
\ifx \showLCCN     \undefined \def \showLCCN      #1{\unskip}     \fi
\ifx \shownote     \undefined \def \shownote      #1{#1}          \fi
\ifx \showarticletitle \undefined \def \showarticletitle #1{#1}   \fi
\ifx \showURL      \undefined \def \showURL       {\relax}        \fi
% invisible to TeX
\providecommand\bibfield[2]{#2}
\providecommand\bibinfo[2]{#2}
\providecommand\natexlab[1]{#1}
\providecommand\showeprint[2][]{arXiv:#2}

\bibitem[Abadi et~al\mbox{.}(2016)]%
        {abadi2016deep}
\bibfield{author}{\bibinfo{person}{Martin Abadi}, \bibinfo{person}{Andy Chu}, \bibinfo{person}{Ian Goodfellow}, \bibinfo{person}{H.~Brendan McMahan}, \bibinfo{person}{Ilya Mironov}, \bibinfo{person}{Kunal Talwar}, {and} \bibinfo{person}{Li Zhang}.} \bibinfo{year}{2016}\natexlab{}.
\newblock \showarticletitle{Deep Learning with Differential Privacy}. In \bibinfo{booktitle}{\emph{Proceedings of the 2016 ACM SIGSAC Conference on Computer and Communications Security}} (Vienna, Austria) \emph{(\bibinfo{series}{CCS '16})}. \bibinfo{publisher}{Association for Computing Machinery}, \bibinfo{address}{New York, NY, USA}, \bibinfo{pages}{308–318}.
\newblock
\showISBNx{9781450341394}
\urldef\tempurl%
\url{https://doi.org/10.1145/2976749.2978318}
\showDOI{\tempurl}


\bibitem[Abowd et~al\mbox{.}(2022)]%
        {abowd20222020}
\bibfield{author}{\bibinfo{person}{John~M Abowd}, \bibinfo{person}{Robert Ashmead}, \bibinfo{person}{Ryan Cumings-Menon}, \bibinfo{person}{Simson Garfinkel}, \bibinfo{person}{Micah Heineck}, \bibinfo{person}{Christine Heiss}, \bibinfo{person}{Robert Johns}, \bibinfo{person}{Daniel Kifer}, \bibinfo{person}{Philip Leclerc}, \bibinfo{person}{Ashwin Machanavajjhala}, {et~al\mbox{.}}} \bibinfo{year}{2022}\natexlab{}.
\newblock \showarticletitle{The 2020 census disclosure avoidance system topdown algorithm}.
\newblock \bibinfo{journal}{\emph{Harvard Data Science Review}}  \bibinfo{volume}{2} (\bibinfo{year}{2022}), \bibinfo{pages}{1--72}.
\newblock


\bibitem[Abowd and Vilhuber(2008)]%
        {abowd2008protective}
\bibfield{author}{\bibinfo{person}{John~M. Abowd} {and} \bibinfo{person}{Lars Vilhuber}.} \bibinfo{year}{2008}\natexlab{}.
\newblock \showarticletitle{How Protective Are Synthetic Data?}. In \bibinfo{booktitle}{\emph{Privacy in Statistical Databases}}, \bibfield{editor}{\bibinfo{person}{Josep Domingo-Ferrer} {and} \bibinfo{person}{Y{\"u}cel Sayg{\i}n}} (Eds.). \bibinfo{publisher}{Springer Berlin Heidelberg}, \bibinfo{address}{Berlin, Heidelberg}, \bibinfo{pages}{239--246}.
\newblock
\showISBNx{978-3-540-87471-3}


\bibitem[Adeleye et~al\mbox{.}(2023)]%
        {adeleye2023publishing}
\bibfield{author}{\bibinfo{person}{Temilola Adeleye}, \bibinfo{person}{Skye Berghel}, \bibinfo{person}{Damien Desfontaines}, \bibinfo{person}{Michael Hay}, \bibinfo{person}{Isaac Johnson}, \bibinfo{person}{Cléo Lemoisson}, \bibinfo{person}{Ashwin Machanavajjhala}, \bibinfo{person}{Tom Magerlein}, \bibinfo{person}{Gabriele Modena}, \bibinfo{person}{David Pujol}, \bibinfo{person}{Daniel Simmons-Marengo}, {and} \bibinfo{person}{Hal Triedman}.} \bibinfo{year}{2023}\natexlab{}.
\newblock \bibinfo{title}{Publishing Wikipedia usage data with strong privacy guarantees}.
\newblock
\newblock
\showeprint[arxiv]{2308.16298}~[cs.CR]
\urldef\tempurl%
\url{https://arxiv.org/abs/2308.16298}
\showURL{%
\tempurl}


\bibitem[Aktay et~al\mbox{.}(2020)]%
        {aktay2020google}
\bibfield{author}{\bibinfo{person}{Ahmet Aktay}, \bibinfo{person}{Shailesh Bavadekar}, \bibinfo{person}{Gwen Cossoul}, \bibinfo{person}{John Davis}, \bibinfo{person}{Damien Desfontaines}, \bibinfo{person}{Alex Fabrikant}, \bibinfo{person}{Evgeniy Gabrilovich}, \bibinfo{person}{Krishna Gadepalli}, \bibinfo{person}{Bryant Gipson}, \bibinfo{person}{Miguel Guevara}, \bibinfo{person}{Chaitanya Kamath}, \bibinfo{person}{Mansi Kansal}, \bibinfo{person}{Ali Lange}, \bibinfo{person}{Chinmoy Mandayam}, \bibinfo{person}{Andrew Oplinger}, \bibinfo{person}{Christopher Pluntke}, \bibinfo{person}{Thomas Roessler}, \bibinfo{person}{Arran Schlosberg}, \bibinfo{person}{Tomer Shekel}, \bibinfo{person}{Swapnil Vispute}, \bibinfo{person}{Mia Vu}, \bibinfo{person}{Gregory Wellenius}, \bibinfo{person}{Brian Williams}, {and} \bibinfo{person}{Royce~J Wilson}.} \bibinfo{year}{2020}\natexlab{}.
\newblock \bibinfo{title}{Google COVID-19 Community Mobility Reports: Anonymization Process Description (version 1.1)}.
\newblock
\newblock
\showeprint[arxiv]{2004.04145}~[cs.CR]
\urldef\tempurl%
\url{https://arxiv.org/abs/2004.04145}
\showURL{%
\tempurl}


\bibitem[Bassolas et~al\mbox{.}(2022)]%
        {bassolas2022reply}
\bibfield{author}{\bibinfo{person}{Aleix Bassolas}, \bibinfo{person}{Hugo Barbosa-Filho}, \bibinfo{person}{Brian Dickinson}, \bibinfo{person}{Xerxes Dotiwalla}, \bibinfo{person}{Paul Eastham}, \bibinfo{person}{Riccardo Gallotti}, \bibinfo{person}{Gourab Ghoshal}, \bibinfo{person}{Bryant Gipson}, \bibinfo{person}{Surendra~A Hazarie}, \bibinfo{person}{Henry Kautz}, {et~al\mbox{.}}} \bibinfo{year}{2022}\natexlab{}.
\newblock \showarticletitle{Reply to: On the difficulty of achieving differential privacy in practice: user-level guarantees in aggregate location data}.
\newblock \bibinfo{journal}{\emph{Nature Communications}} \bibinfo{volume}{13}, \bibinfo{number}{1} (\bibinfo{year}{2022}), \bibinfo{pages}{30}.
\newblock


\bibitem[Bun and Steinke(2016)]%
        {bun2016concentrated}
\bibfield{author}{\bibinfo{person}{Mark Bun} {and} \bibinfo{person}{Thomas Steinke}.} \bibinfo{year}{2016}\natexlab{}.
\newblock \showarticletitle{Concentrated Differential Privacy: Simplifications, Extensions, and Lower Bounds}. In \bibinfo{booktitle}{\emph{Theory of Cryptography}}, \bibfield{editor}{\bibinfo{person}{Martin Hirt} {and} \bibinfo{person}{Adam Smith}} (Eds.). \bibinfo{publisher}{Springer Berlin Heidelberg}, \bibinfo{address}{Berlin, Heidelberg}, \bibinfo{pages}{635--658}.
\newblock
\showISBNx{978-3-662-53641-4}


\bibitem[Canonne et~al\mbox{.}(2020)]%
        {canonne2020discrete}
\bibfield{author}{\bibinfo{person}{Cl{\'e}ment~L Canonne}, \bibinfo{person}{Gautam Kamath}, {and} \bibinfo{person}{Thomas Steinke}.} \bibinfo{year}{2020}\natexlab{}.
\newblock \showarticletitle{The discrete gaussian for differential privacy}.
\newblock \bibinfo{journal}{\emph{Advances in Neural Information Processing Systems}}  \bibinfo{volume}{33} (\bibinfo{year}{2020}), \bibinfo{pages}{15676--15688}.
\newblock


\bibitem[Cummings et~al\mbox{.}(2024)]%
        {cummings2024attaxonomy}
\bibfield{author}{\bibinfo{person}{Rachel Cummings}, \bibinfo{person}{Shlomi Hod}, \bibinfo{person}{Jayshree Sarathy}, {and} \bibinfo{person}{Marika Swanberg}.} \bibinfo{year}{2024}\natexlab{}.
\newblock \bibinfo{title}{ATTAXONOMY: Unpacking Differential Privacy Guarantees Against Practical Adversaries}.
\newblock
\newblock
\showeprint[arxiv]{2405.01716}~[cs.CR]
\urldef\tempurl%
\url{https://arxiv.org/abs/2405.01716}
\showURL{%
\tempurl}


\bibitem[Cummings et~al\mbox{.}(2021)]%
        {cummings2021need}
\bibfield{author}{\bibinfo{person}{Rachel Cummings}, \bibinfo{person}{Gabriel Kaptchuk}, {and} \bibinfo{person}{Elissa~M. Redmiles}.} \bibinfo{year}{2021}\natexlab{}.
\newblock \showarticletitle{"I need a better description": An Investigation Into User Expectations For Differential Privacy}. In \bibinfo{booktitle}{\emph{Proceedings of the 2021 ACM SIGSAC Conference on Computer and Communications Security}} (Virtual Event, Republic of Korea) \emph{(\bibinfo{series}{CCS '21})}. \bibinfo{publisher}{Association for Computing Machinery}, \bibinfo{address}{New York, NY, USA}, \bibinfo{pages}{3037–3052}.
\newblock
\showISBNx{9781450384544}
\urldef\tempurl%
\url{https://doi.org/10.1145/3460120.3485252}
\showDOI{\tempurl}


\bibitem[Dibia et~al\mbox{.}(2024)]%
        {dibia2024sok}
\bibfield{author}{\bibinfo{person}{Onyinye Dibia}, \bibinfo{person}{Brad Stenger}, \bibinfo{person}{Steven Baldasty}, \bibinfo{person}{Mako Bates}, \bibinfo{person}{Ivoline~C. Ngong}, \bibinfo{person}{Yuanyuan Feng}, {and} \bibinfo{person}{Joseph~P. Near}.} \bibinfo{year}{2024}\natexlab{}.
\newblock \bibinfo{title}{SoK: Usability Studies in Differential Privacy}.
\newblock
\newblock
\showeprint[arxiv]{2412.16825}~[cs.HC]
\urldef\tempurl%
\url{https://arxiv.org/abs/2412.16825}
\showURL{%
\tempurl}


\bibitem[Duncan and Lambert(1986)]%
        {duncan1986disclosure}
\bibfield{author}{\bibinfo{person}{George~T Duncan} {and} \bibinfo{person}{Diane Lambert}.} \bibinfo{year}{1986}\natexlab{}.
\newblock \showarticletitle{Disclosure-limited data dissemination}.
\newblock \bibinfo{journal}{\emph{Journal of the American statistical association}} \bibinfo{volume}{81}, \bibinfo{number}{393} (\bibinfo{year}{1986}), \bibinfo{pages}{10--18}.
\newblock


\bibitem[Dwork et~al\mbox{.}(2006a)]%
        {dwork2006our}
\bibfield{author}{\bibinfo{person}{Cynthia Dwork}, \bibinfo{person}{Krishnaram Kenthapadi}, \bibinfo{person}{Frank McSherry}, \bibinfo{person}{Ilya Mironov}, {and} \bibinfo{person}{Moni Naor}.} \bibinfo{year}{2006}\natexlab{a}.
\newblock \showarticletitle{Our Data, Ourselves: Privacy Via Distributed Noise Generation}. In \bibinfo{booktitle}{\emph{Advances in Cryptology - EUROCRYPT 2006}}, \bibfield{editor}{\bibinfo{person}{Serge Vaudenay}} (Ed.). \bibinfo{publisher}{Springer Berlin Heidelberg}, \bibinfo{address}{Berlin, Heidelberg}, \bibinfo{pages}{486--503}.
\newblock
\showISBNx{978-3-540-34547-3}


\bibitem[Dwork et~al\mbox{.}(2006b)]%
        {dwork2006calibrating}
\bibfield{author}{\bibinfo{person}{Cynthia Dwork}, \bibinfo{person}{Frank McSherry}, \bibinfo{person}{Kobbi Nissim}, {and} \bibinfo{person}{Adam Smith}.} \bibinfo{year}{2006}\natexlab{b}.
\newblock \showarticletitle{Calibrating Noise to Sensitivity in Private Data Analysis}. In \bibinfo{booktitle}{\emph{Theory of Cryptography}}, \bibfield{editor}{\bibinfo{person}{Shai Halevi} {and} \bibinfo{person}{Tal Rabin}} (Eds.). \bibinfo{publisher}{Springer Berlin Heidelberg}, \bibinfo{address}{Berlin, Heidelberg}, \bibinfo{pages}{265--284}.
\newblock
\showISBNx{978-3-540-32732-5}


\bibitem[Dwork and Pottenger(2013)]%
        {dwork2013toward}
\bibfield{author}{\bibinfo{person}{Cynthia Dwork} {and} \bibinfo{person}{Rebecca Pottenger}.} \bibinfo{year}{2013}\natexlab{}.
\newblock \showarticletitle{Toward practicing privacy}.
\newblock \bibinfo{journal}{\emph{Journal of the American Medical Informatics Association}} \bibinfo{volume}{20}, \bibinfo{number}{1} (\bibinfo{year}{2013}), \bibinfo{pages}{102--108}.
\newblock


\bibitem[Dwork et~al\mbox{.}(2014)]%
        {dwork2014algorithmic}
\bibfield{author}{\bibinfo{person}{Cynthia Dwork}, \bibinfo{person}{Aaron Roth}, {et~al\mbox{.}}} \bibinfo{year}{2014}\natexlab{}.
\newblock \showarticletitle{The algorithmic foundations of differential privacy}.
\newblock \bibinfo{journal}{\emph{Foundations and Trends{\textregistered} in Theoretical Computer Science}} \bibinfo{volume}{9}, \bibinfo{number}{3--4} (\bibinfo{year}{2014}), \bibinfo{pages}{211--407}.
\newblock


\bibitem[Dwork et~al\mbox{.}(2010)]%
        {dwork2010boosting}
\bibfield{author}{\bibinfo{person}{Cynthia Dwork}, \bibinfo{person}{Guy~N. Rothblum}, {and} \bibinfo{person}{Salil Vadhan}.} \bibinfo{year}{2010}\natexlab{}.
\newblock \showarticletitle{Boosting and Differential Privacy}. In \bibinfo{booktitle}{\emph{2010 IEEE 51st Annual Symposium on Foundations of Computer Science}}. \bibinfo{publisher}{IEEE}, \bibinfo{address}{Piscataway, NJ, USA}, \bibinfo{pages}{51--60}.
\newblock
\urldef\tempurl%
\url{https://doi.org/10.1109/FOCS.2010.12}
\showDOI{\tempurl}


\bibitem[Franzen et~al\mbox{.}(2022)]%
        {franzen2022private}
\bibfield{author}{\bibinfo{person}{Daniel Franzen}, \bibinfo{person}{Saskia Nu\~{n}ez~von Voigt}, \bibinfo{person}{Peter S\"{o}rries}, \bibinfo{person}{Florian Tschorsch}, {and} \bibinfo{person}{Claudia M\"{u}ller-Birn}.} \bibinfo{year}{2022}\natexlab{}.
\newblock \showarticletitle{Am I Private and If So, how Many? Communicating Privacy Guarantees of Differential Privacy with Risk Communication Formats}. In \bibinfo{booktitle}{\emph{Proceedings of the 2022 ACM SIGSAC Conference on Computer and Communications Security}} (Los Angeles, CA, USA) \emph{(\bibinfo{series}{CCS '22})}. \bibinfo{publisher}{Association for Computing Machinery}, \bibinfo{address}{New York, NY, USA}, \bibinfo{pages}{1125–1139}.
\newblock
\showISBNx{9781450394505}
\urldef\tempurl%
\url{https://doi.org/10.1145/3548606.3560693}
\showDOI{\tempurl}


\bibitem[Gaboardi et~al\mbox{.}(2020)]%
        {gaboardi2020programming}
\bibfield{author}{\bibinfo{person}{Marco Gaboardi}, \bibinfo{person}{Michael Hay}, {and} \bibinfo{person}{Salil Vadhan}.} \bibinfo{year}{2020}\natexlab{}.
\newblock \bibinfo{title}{A programming framework for opendp}.
\newblock
\newblock


\bibitem[Herda{\u{g}}delen et~al\mbox{.}(2020)]%
        {herdaugdelen2020protecting}
\bibfield{author}{\bibinfo{person}{A Herda{\u{g}}delen}, \bibinfo{person}{Alex Dow}, \bibinfo{person}{S Bogdan}, \bibinfo{person}{M Payman}, {and} \bibinfo{person}{A Pompe}.} \bibinfo{year}{2020}\natexlab{}.
\newblock \bibinfo{title}{Protecting privacy in Facebook mobility data during the COVID-19 response}.
\newblock
\newblock


\bibitem[Hotz et~al\mbox{.}(2022)]%
        {hotz2022balancing}
\bibfield{author}{\bibinfo{person}{V~Joseph Hotz}, \bibinfo{person}{Christopher~R Bollinger}, \bibinfo{person}{Tatiana Komarova}, \bibinfo{person}{Charles~F Manski}, \bibinfo{person}{Robert~A Moffitt}, \bibinfo{person}{Denis Nekipelov}, \bibinfo{person}{Aaron Sojourner}, {and} \bibinfo{person}{Bruce~D Spencer}.} \bibinfo{year}{2022}\natexlab{}.
\newblock \showarticletitle{Balancing data privacy and usability in the federal statistical system}.
\newblock \bibinfo{journal}{\emph{Proceedings of the National Academy of Sciences}} \bibinfo{volume}{119}, \bibinfo{number}{31} (\bibinfo{year}{2022}), \bibinfo{pages}{e2104906119}.
\newblock


\bibitem[John et~al\mbox{.}(2021)]%
        {john2021decision}
\bibfield{author}{\bibinfo{person}{Mark F.~St. John}, \bibinfo{person}{Grit Denker}, \bibinfo{person}{Peeter Laud}, \bibinfo{person}{Karsten Martiny}, \bibinfo{person}{Alisa Pankova}, {and} \bibinfo{person}{Dusko Pavlovic}.} \bibinfo{year}{2021}\natexlab{}.
\newblock \showarticletitle{Decision Support for Sharing Data using Differential Privacy}. In \bibinfo{booktitle}{\emph{2021 IEEE Symposium on Visualization for Cyber Security (VizSec)}}. \bibinfo{publisher}{IEEE}, \bibinfo{address}{Piscataway, NJ, USA}, \bibinfo{pages}{26--35}.
\newblock
\urldef\tempurl%
\url{https://doi.org/10.1109/VizSec53666.2021.00008}
\showDOI{\tempurl}


\bibitem[Kairouz et~al\mbox{.}(2015)]%
        {kairouz2015composition}
\bibfield{author}{\bibinfo{person}{Peter Kairouz}, \bibinfo{person}{Sewoong Oh}, {and} \bibinfo{person}{Pramod Viswanath}.} \bibinfo{year}{2015}\natexlab{}.
\newblock \showarticletitle{The Composition Theorem for Differential Privacy}. In \bibinfo{booktitle}{\emph{Proceedings of the 32nd International Conference on Machine Learning}} \emph{(\bibinfo{series}{Proceedings of Machine Learning Research}, Vol.~\bibinfo{volume}{37})}, \bibfield{editor}{\bibinfo{person}{Francis Bach} {and} \bibinfo{person}{David Blei}} (Eds.). \bibinfo{publisher}{PMLR}, \bibinfo{address}{Lille, France}, \bibinfo{pages}{1376--1385}.
\newblock
\urldef\tempurl%
\url{https://proceedings.mlr.press/v37/kairouz15.html}
\showURL{%
\tempurl}


\bibitem[Kasiviswanathan and Smith(2014)]%
        {kasiviswanathan2014semantics}
\bibfield{author}{\bibinfo{person}{Shiva~P. Kasiviswanathan} {and} \bibinfo{person}{Adam Smith}.} \bibinfo{year}{2014}\natexlab{}.
\newblock \showarticletitle{On the ’Semantics’ of Differential Privacy: A Bayesian Formulation}.
\newblock \bibinfo{journal}{\emph{Journal of Privacy and Confidentiality}} \bibinfo{volume}{6}, \bibinfo{number}{1} (\bibinfo{date}{Jun.} \bibinfo{year}{2014}), \bibinfo{pages}{1--16}.
\newblock
\urldef\tempurl%
\url{https://doi.org/10.29012/jpc.v6i1.634}
\showDOI{\tempurl}


\bibitem[Kazan and Reiter(2025)]%
        {kazan2025assessing}
\bibfield{author}{\bibinfo{person}{Zeki Kazan} {and} \bibinfo{person}{Jerome Reiter}.} \bibinfo{year}{2025}\natexlab{}.
\newblock \showarticletitle{Assessing Statistical Disclosure Risk for Differentially Private, Hierarchical Count Data, with Application to the 2020 US Decennial Census}.
\newblock \bibinfo{journal}{\emph{Statistica Sinica}}  \bibinfo{volume}{35} (\bibinfo{year}{2025}), \bibinfo{pages}{1--21}.
\newblock


\bibitem[Kazan and Reiter(2024)]%
        {kazan2024prior}
\bibfield{author}{\bibinfo{person}{Zeki Kazan} {and} \bibinfo{person}{Jerome~P. Reiter}.} \bibinfo{year}{2024}\natexlab{}.
\newblock \showarticletitle{Prior-itizing Privacy: A Bayesian Approach to Setting the Privacy Budget in Differential Privacy}.
\newblock \bibinfo{journal}{\emph{Advances in Neural Information Processing Systems}}  \bibinfo{volume}{37} (\bibinfo{year}{2024}), \bibinfo{pages}{90384--90430}.
\newblock


\bibitem[Kifer et~al\mbox{.}(2022)]%
        {kifer2022bayesian}
\bibfield{author}{\bibinfo{person}{Daniel Kifer}, \bibinfo{person}{John~M. Abowd}, \bibinfo{person}{Robert Ashmead}, \bibinfo{person}{Ryan Cumings-Menon}, \bibinfo{person}{Philip Leclerc}, \bibinfo{person}{Ashwin Machanavajjhala}, \bibinfo{person}{William Sexton}, {and} \bibinfo{person}{Pavel Zhuravlev}.} \bibinfo{year}{2022}\natexlab{}.
\newblock \bibinfo{title}{Bayesian and Frequentist Semantics for Common Variations of Differential Privacy: Applications to the 2020 Census}.
\newblock
\newblock
\showeprint[arxiv]{2209.03310}~[cs.CR]
\urldef\tempurl%
\url{https://arxiv.org/abs/2209.03310}
\showURL{%
\tempurl}


\bibitem[Kifer and Lin(2012)]%
        {kifer2012axiomatic}
\bibfield{author}{\bibinfo{person}{Daniel Kifer} {and} \bibinfo{person}{Bing-Rong Lin}.} \bibinfo{year}{2012}\natexlab{}.
\newblock \showarticletitle{An Axiomatic View of Statistical Privacy and Utility}.
\newblock \bibinfo{journal}{\emph{Journal of Privacy and Confidentiality}} \bibinfo{volume}{4}, \bibinfo{number}{1} (\bibinfo{date}{Jul.} \bibinfo{year}{2012}), \bibinfo{pages}{5–--49}.
\newblock
\urldef\tempurl%
\url{https://doi.org/10.29012/jpc.v4i1.610}
\showDOI{\tempurl}


\bibitem[Kifer and Machanavajjhala(2014)]%
        {kifer2014pufferfish}
\bibfield{author}{\bibinfo{person}{Daniel Kifer} {and} \bibinfo{person}{Ashwin Machanavajjhala}.} \bibinfo{year}{2014}\natexlab{}.
\newblock \showarticletitle{Pufferfish: A framework for mathematical privacy definitions}.
\newblock \bibinfo{journal}{\emph{ACM Transactions on Database Systems (TODS)}} \bibinfo{volume}{39}, \bibinfo{number}{1} (\bibinfo{year}{2014}), \bibinfo{pages}{1--36}.
\newblock


\bibitem[Labs(2023)]%
        {TumultLabs_DeptEd}
\bibfield{author}{\bibinfo{person}{Tumult Labs}.} \bibinfo{year}{2023}\natexlab{}.
\newblock \bibinfo{title}{Illuminating College Outcomes | Tumult Labs}.
\newblock
\newblock
\urldef\tempurl%
\url{https://www.tmlt.io/casestudy/illuminating-college-outcomes-while-protecting-privacy}
\showURL{%
\tempurl}


\bibitem[Lee and Clifton(2011)]%
        {lee2011much}
\bibfield{author}{\bibinfo{person}{Jaewoo Lee} {and} \bibinfo{person}{Chris Clifton}.} \bibinfo{year}{2011}\natexlab{}.
\newblock \showarticletitle{How Much Is Enough? Choosing $\epsilon$ for Differential Privacy}. In \bibinfo{booktitle}{\emph{Information Security}}, \bibfield{editor}{\bibinfo{person}{Xuejia Lai}, \bibinfo{person}{Jianying Zhou}, {and} \bibinfo{person}{Hui Li}} (Eds.). \bibinfo{publisher}{Springer Berlin Heidelberg}, \bibinfo{address}{Berlin, Heidelberg}, \bibinfo{pages}{325--340}.
\newblock
\showISBNx{978-3-642-24861-0}


\bibitem[Machanavajjhala et~al\mbox{.}(2008)]%
        {machanavajjhala2008privacy}
\bibfield{author}{\bibinfo{person}{Ashwin Machanavajjhala}, \bibinfo{person}{Daniel Kifer}, \bibinfo{person}{John Abowd}, \bibinfo{person}{Johannes Gehrke}, {and} \bibinfo{person}{Lars Vilhuber}.} \bibinfo{year}{2008}\natexlab{}.
\newblock \showarticletitle{Privacy: Theory meets Practice on the Map}. In \bibinfo{booktitle}{\emph{2008 IEEE 24th International Conference on Data Engineering}}. \bibinfo{publisher}{IEEE}, \bibinfo{address}{IEEE}, \bibinfo{pages}{277--286}.
\newblock
\urldef\tempurl%
\url{https://doi.org/10.1109/ICDE.2008.4497436}
\showDOI{\tempurl}


\bibitem[McClure and Reiter(2012)]%
        {mcclure2012differential}
\bibfield{author}{\bibinfo{person}{David McClure} {and} \bibinfo{person}{Jerome~P Reiter}.} \bibinfo{year}{2012}\natexlab{}.
\newblock \showarticletitle{Differential Privacy and Statistical Disclosure Risk Measures: An Investigation with Binary Synthetic Data.}
\newblock \bibinfo{journal}{\emph{Transactions on Data Privacy}} \bibinfo{volume}{5}, \bibinfo{number}{3} (\bibinfo{year}{2012}), \bibinfo{pages}{535--552}.
\newblock


\bibitem[Meiser(2018)]%
        {meiser2018approximate}
\bibfield{author}{\bibinfo{person}{Sebastian Meiser}.} \bibinfo{year}{2018}\natexlab{}.
\newblock \bibinfo{title}{Approximate and probabilistic differential privacy definitions}.
\newblock
\newblock


\bibitem[Messing et~al\mbox{.}(2020)]%
        {DVN/TDOAPG_2020}
\bibfield{author}{\bibinfo{person}{Solomon Messing}, \bibinfo{person}{Christina DeGregorio}, \bibinfo{person}{Bennett Hillenbrand}, \bibinfo{person}{Gary King}, \bibinfo{person}{Saurav Mahanti}, \bibinfo{person}{Zagreb Mukerjee}, \bibinfo{person}{Chaya Nayak}, \bibinfo{person}{Nate Persily}, \bibinfo{person}{Bogdan State}, {and} \bibinfo{person}{Arjun Wilkins}.} \bibinfo{year}{2020}\natexlab{}.
\newblock \bibinfo{title}{{Facebook Privacy-Protected Full URLs Data Set}}.
\newblock
\newblock
\urldef\tempurl%
\url{https://doi.org/10.7910/DVN/TDOAPG}
\showDOI{\tempurl}


\bibitem[Murtagh and Vadhan(2016)]%
        {murtagh2015complexity}
\bibfield{author}{\bibinfo{person}{Jack Murtagh} {and} \bibinfo{person}{Salil Vadhan}.} \bibinfo{year}{2016}\natexlab{}.
\newblock \showarticletitle{The Complexity of Computing the Optimal Composition of Differential Privacy}. In \bibinfo{booktitle}{\emph{Theory of Cryptography}}, \bibfield{editor}{\bibinfo{person}{Eyal Kushilevitz} {and} \bibinfo{person}{Tal Malkin}} (Eds.). \bibinfo{publisher}{Springer Berlin Heidelberg}, \bibinfo{address}{Berlin, Heidelberg}, \bibinfo{pages}{157--175}.
\newblock
\showISBNx{978-3-662-49096-9}


\bibitem[Nanayakkara et~al\mbox{.}(2023)]%
        {nanayakkara2023chances}
\bibfield{author}{\bibinfo{person}{Priyanka Nanayakkara}, \bibinfo{person}{Mary~Anne Smart}, \bibinfo{person}{Rachel Cummings}, \bibinfo{person}{Gabriel Kaptchuk}, {and} \bibinfo{person}{Elissa~M. Redmiles}.} \bibinfo{year}{2023}\natexlab{}.
\newblock \showarticletitle{What Are the Chances? Explaining the Epsilon Parameter in Differential Privacy}. In \bibinfo{booktitle}{\emph{32nd USENIX Security Symposium (USENIX Security 23)}}. \bibinfo{publisher}{USENIX Association}, \bibinfo{address}{Anaheim, CA}, \bibinfo{pages}{1613--1630}.
\newblock
\showISBNx{978-1-939133-37-3}
\urldef\tempurl%
\url{https://www.usenix.org/conference/usenixsecurity23/presentation/nanayakkara}
\showURL{%
\tempurl}


\bibitem[Pankova and Laud(2022)]%
        {pankova2022interpreting}
\bibfield{author}{\bibinfo{person}{Alisa Pankova} {and} \bibinfo{person}{Peeter Laud}.} \bibinfo{year}{2022}\natexlab{}.
\newblock \showarticletitle{Interpreting Epsilon of Differential Privacy in Terms of Advantage in Guessing or Approximating Sensitive Attributes}. In \bibinfo{booktitle}{\emph{2022 IEEE 35th Computer Security Foundations Symposium (CSF)}}. \bibinfo{publisher}{IEEE}, \bibinfo{address}{Piscataway, NJ, USA}, \bibinfo{pages}{96--111}.
\newblock
\urldef\tempurl%
\url{https://doi.org/10.1109/CSF54842.2022.9919656}
\showDOI{\tempurl}


\bibitem[Pereira et~al\mbox{.}(2021)]%
        {pereira2021us}
\bibfield{author}{\bibinfo{person}{Mayana Pereira}, \bibinfo{person}{Allen Kim}, \bibinfo{person}{Joshua Allen}, \bibinfo{person}{Kevin White}, \bibinfo{person}{Juan~Lavista Ferres}, {and} \bibinfo{person}{Rahul Dodhia}.} \bibinfo{year}{2021}\natexlab{}.
\newblock \bibinfo{title}{U.S. Broadband Coverage Data Set: A Differentially Private Data Release}.
\newblock
\newblock
\urldef\tempurl%
\url{https://europepmc.org/article/PPR/PPR343016}
\showURL{%
\tempurl}


\bibitem[Spectus(2022)]%
        {Spectus}
\bibfield{author}{\bibinfo{person}{Spectus}.} \bibinfo{year}{2022}\natexlab{}.
\newblock \bibinfo{title}{Differential Privacy}.
\newblock
\newblock
\urldef\tempurl%
\url{https://spectus.ai/wp-content/uploads/2022/10/Spectus_DPWhitepaper_v01b.pdf}
\showURL{%
\tempurl}


\bibitem[Steinke(2022)]%
        {steinke2022composition}
\bibfield{author}{\bibinfo{person}{Thomas Steinke}.} \bibinfo{year}{2022}\natexlab{}.
\newblock \bibinfo{title}{Composition of Differential Privacy \& Privacy Amplification by Subsampling}.
\newblock
\newblock
\showeprint[arxiv]{2210.00597}~[cs.CR]
\urldef\tempurl%
\url{https://arxiv.org/abs/2210.00597}
\showURL{%
\tempurl}


\bibitem[Tang et~al\mbox{.}(2017)]%
        {tang2017privacy}
\bibfield{author}{\bibinfo{person}{Jun Tang}, \bibinfo{person}{Aleksandra Korolova}, \bibinfo{person}{Xiaolong Bai}, \bibinfo{person}{Xueqiang Wang}, {and} \bibinfo{person}{Xiaofeng Wang}.} \bibinfo{year}{2017}\natexlab{}.
\newblock \bibinfo{title}{Privacy Loss in Apple's Implementation of Differential Privacy on MacOS 10.12}.
\newblock
\newblock
\showeprint[arxiv]{1709.02753}~[cs.CR]
\urldef\tempurl%
\url{https://arxiv.org/abs/1709.02753}
\showURL{%
\tempurl}


\bibitem[Thudi et~al\mbox{.}(2024)]%
        {thudi2024differential}
\bibfield{author}{\bibinfo{person}{Anvith Thudi}, \bibinfo{person}{Ilia Shumailov}, \bibinfo{person}{Franziska Boenisch}, {and} \bibinfo{person}{Nicolas Papernot}.} \bibinfo{year}{2024}\natexlab{}.
\newblock \bibinfo{title}{From Differential Privacy to Bounds on Membership Inference: Less can be More}.
\newblock
\newblock
\showISSN{2835-8856}
\urldef\tempurl%
\url{https://openreview.net/forum?id=daXqjb6dVE}
\showURL{%
\tempurl}


\bibitem[Wood et~al\mbox{.}(2018)]%
        {wood2018differential}
\bibfield{author}{\bibinfo{person}{Alexandra Wood}, \bibinfo{person}{Micah Altman}, \bibinfo{person}{Aaron Bembenek}, \bibinfo{person}{Mark Bun}, \bibinfo{person}{Marco Gaboardi}, \bibinfo{person}{James Honaker}, \bibinfo{person}{Kobbi Nissim}, \bibinfo{person}{David~R O'Brien}, \bibinfo{person}{Thomas Steinke}, {and} \bibinfo{person}{Salil Vadhan}.} \bibinfo{year}{2018}\natexlab{}.
\newblock \showarticletitle{Differential privacy: A primer for a non-technical audience}.
\newblock \bibinfo{journal}{\emph{Vand. J. Ent. \& Tech. L.}}  \bibinfo{volume}{21} (\bibinfo{year}{2018}), \bibinfo{pages}{209}.
\newblock


\bibitem[Wood et~al\mbox{.}(2020)]%
        {wood2020designing}
\bibfield{author}{\bibinfo{person}{Alexandra Wood}, \bibinfo{person}{Micah Altman}, \bibinfo{person}{Kobbi Nissim}, {and} \bibinfo{person}{Salil Vadhan}.} \bibinfo{year}{2020}\natexlab{}.
\newblock \bibinfo{title}{Designing access with differential privacy}.
\newblock
\newblock


\bibitem[Zhao et~al\mbox{.}(2019)]%
        {zhao2019reviewing}
\bibfield{author}{\bibinfo{person}{Jun Zhao}, \bibinfo{person}{Teng Wang}, \bibinfo{person}{Tao Bai}, \bibinfo{person}{Kwok-Yan Lam}, \bibinfo{person}{Zhiying Xu}, \bibinfo{person}{Shuyu Shi}, \bibinfo{person}{Xuebin Ren}, \bibinfo{person}{Xinyu Yang}, \bibinfo{person}{Yang Liu}, {and} \bibinfo{person}{Han Yu}.} \bibinfo{year}{2019}\natexlab{}.
\newblock \bibinfo{title}{Reviewing and Improving the Gaussian Mechanism for Differential Privacy}.
\newblock
\newblock
\showeprint[arxiv]{1911.12060}~[cs.CR]
\urldef\tempurl%
\url{https://arxiv.org/abs/1911.12060}
\showURL{%
\tempurl}


\bibitem[Zhu et~al\mbox{.}(2022)]%
        {zhu2022optimal}
\bibfield{author}{\bibinfo{person}{Yuqing Zhu}, \bibinfo{person}{Jinshuo Dong}, {and} \bibinfo{person}{Yu-Xiang Wang}.} \bibinfo{year}{2022}\natexlab{}.
\newblock \showarticletitle{Optimal Accounting of Differential Privacy via Characteristic Function}. In \bibinfo{booktitle}{\emph{Proceedings of The 25th International Conference on Artificial Intelligence and Statistics}} \emph{(\bibinfo{series}{Proceedings of Machine Learning Research}, Vol.~\bibinfo{volume}{151})}, \bibfield{editor}{\bibinfo{person}{Gustau Camps-Valls}, \bibinfo{person}{Francisco J.~R. Ruiz}, {and} \bibinfo{person}{Isabel Valera}} (Eds.). \bibinfo{publisher}{PMLR}, \bibinfo{address}{PMLR}, \bibinfo{pages}{4782--4817}.
\newblock


\end{thebibliography}

\appendix

\section{Omitted Proofs} \label{sec:proofs}

In this section, we prove the results presented in Section \ref{sec:results}. 

\subsection{Proof of Theorem \ref{thm:PLRV_posterior}} \label{proof:PLRV_posterior}

We begin with Theorem \ref{thm:PLRV_posterior}, which proceeds by decomposing the adversary's posterior probability via Bayes' Theorem and recognizing the form of a privacy loss random variable. The result is restated below.

\begin{reptheorem}{thm:PLRV_posterior}
    For the setting described in Section \ref{sec:setting}, define the privacy loss random variables $Z_i = \PL(M(\data) ~\Vert~ M(\data_{-i}))$ and $Z_i' = \PL(M(\data_{-i}) ~\Vert~ M(\data))$. Then 
    \begin{align} \tag{\ref{eq:posterior_PLRV}}
        X_i = 
        \begin{cases}
            \frac{p_i}{p_i + (1-p_i)e^{-Z_i}}, & \mbox{if } Y \leftarrow M(\data); \\
            \frac{p_i}{p_i + (1-p_i)e^{Z_i'}}, & \mbox{if } Y \leftarrow M(\data_{-i}).
        \end{cases}
    \end{align}
\end{reptheorem}

\begin{proof}
    Consider the form of the function $f_i$. By Bayes' Theorem,
    \begin{align*}
       f_i(y, p_i) &= P_\M[I_i = 1 \mid Y = y] \\
       &= \frac{P_\M[Y = y \mid I_i = 1] \, p_i}{P_\M[Y = y \mid I_i = 1] \, p_i + P_\M[Y = y \mid I_i = 0] (1-p_i)} \\
       &= \frac{p_i}{p_i + \frac{P_\M[Y = y \mid I_i = 0]}{P_\M[Y = y \mid I_i = 1]} (1-p_i)}.
    \end{align*}
    Since the adversary's prior knowledge $\M$ includes complete knowledge of $\data_{-i}$, under $\M$, conditioning on $I_i = 0$ implies $Y \leftarrow M(\data_{-i})$. Thus,
    \begin{align*}
        P_\M[Y = y \mid I_i = 0] = P[M(\data_{-i}) = y].
    \end{align*}
    Since the adversary's prior knowledge $\M$ additionally includes the attributes of person $i$, conditioning on $I_i = 1$ implies $Y \leftarrow M(\data)$. Thus,
    \begin{align*}
        P_\M[Y = y \mid I_i = 1] = P[M(\data) = y].
    \end{align*}
    The expression of interest can now be rewritten as follows.
    \begin{align}
       f_i(y, p_i) &= \frac{p_i}{p_i + \frac{P[M(\data_{-i}) = y]}{P[M(\data) = y]} (1-p_i)} \nonumber \\
       &= \frac{p_i}{p_i + \exp\left\{\log\left(\frac{P[M(\data_{-i}) = y]}{P[M(\data) = y]} \right)\right\} (1-p_i)} \label{eq:f_equal_PLRV2} \\
       &= \frac{p_i}{p_i + \exp\left\{-\log\left(\frac{P[M(\data) = y]}{P[M(\data_{-i}) = y]} \right)\right\} (1-p_i)}. \label{eq:f_equal_PLRV}
   \end{align}
   If $Y \leftarrow M(\data)$, then by (\ref{eq:f_equal_PLRV}), 
   \begin{align*}
       X_i = f_i(Y, p_i) = \frac{p_i}{p_i + (1-p_i)e^{-Z_i}}.
   \end{align*}
   If $Y \leftarrow M(\data_{-i})$, then by (\ref{eq:f_equal_PLRV2}),
   \begin{align*}
       X_i = f_i(Y, p_i) = \frac{p_i}{p_i + (1-p_i)e^{Z_i'}}.
   \end{align*}
   The result follows.
\end{proof}

\subsection{Proof of Theorem \ref{thm:PDP_posterior}} \label{proof:PDP_posterior}

We next prove Theorem \ref{thm:PDP_posterior}. The proof proceeds by using Theorem \ref{thm:PLRV_posterior} to transform probabilistic statements about $X_i$ into probabilistic statements about privacy loss random variables and vice versa. The result is restated below.

\begin{reptheorem}{thm:PDP_posterior}
    $M$ satisfies $(\varepsilon, \delta)$-PDP if and only if in the setting described in Section \ref{sec:setting}, for any database $\data$, target $i$, $p_i \in [0,1]$, and under both $Y \leftarrow M(\data)$ and $Y \leftarrow M(\data_{-i})$, the distribution of $X_i$ satisfies
    \begin{align} \tag{\ref{eq:eps_delta_post}}
        P\left[\frac{p_i}{p_i + (1-p_i)e^{\varepsilon}} \leq X_i \leq \frac{p_i}{p_i + (1-p_i)e^{-\varepsilon}} \right] \geq 1-\delta.
    \end{align}
\end{reptheorem}

\begin{proof}
   ($\Longrightarrow$) Suppose the mechanism $M$ satisfies $(\varepsilon, \delta)$-PDP. Let $\data$, $i$, and $p_i$ be arbitrary and define the random variables $Z_i = \PL(M(\data) ~\Vert~ M(\data_{-i}))$ and $Z_i' = \PL(M(\data_{-i}) ~\Vert~ M(\data))$. By Theorem \ref{thm:PLRV_posterior}, if $Y \leftarrow M(\data)$, then
    \begin{align*}
       1-&\delta \leq P[-\varepsilon \leq Z_i \leq \varepsilon] \\
       &= P\left[\frac{p_i}{p_i + (1-p_i)e^{\varepsilon}} \leq \frac{p_i}{p_i + (1-p_i)e^{-Z_i}} \leq \frac{p_i}{p_i + (1-p_i)e^{-\varepsilon}} \right] \\
       &= P\left[\frac{p_i}{p_i + (1-p_i)e^{\varepsilon}} \leq X_i \leq \frac{p_i}{p_i + (1-p_i)e^{-\varepsilon}} \right].
    \end{align*}
    If $Y \leftarrow M(\data_{-i})$, then
    \begin{align*}
       1-&\delta \leq P[-\varepsilon \leq Z_i' \leq \varepsilon] \\
       &= P\left[\frac{p_i}{p_i + (1-p_i)e^{\varepsilon}} \leq \frac{p_i}{p_i + (1-p_i)e^{Z_i'}} \leq \frac{p_i}{p_i + (1-p_i)e^{-\varepsilon}} \right] \\
       &= P\left[\frac{p_i}{p_i + (1-p_i)e^{\varepsilon}} \leq X_i \leq \frac{p_i}{p_i + (1-p_i)e^{-\varepsilon}} \right].
    \end{align*}
    Thus, in either case (\ref{eq:eps_delta_post}) is satisfied. 

   ($\Longleftarrow$) Now suppose (\ref{eq:eps_delta_post}) is satisfied for all $\data$, all $i$, and all $p_i \in [0,1]$, under both $Y \leftarrow M(\data)$ and $Y \leftarrow M(\data_{-i})$. Let $\data_1, \data_2$ be an arbitrary pair of neighboring databases and let $Z = \PL(M(\data_1) ~\Vert~ M(\data_2))$.

   Suppose $\data_1$ has one more row than $\data_2$ and let $i$ index the observation that is in $\data_1$, but not $\data_2$. By Theorem \ref{thm:PLRV_posterior} with $\data = \data_1$, it follows that if $Y \leftarrow M(\data_1)$, then $X_i = p_i/(p_i + (1-p_i)e^{-Z})$. Thus, from (\ref{eq:eps_delta_post}) with $Y \leftarrow M(\data_1)$,
   \begin{align*}
       1-&\delta \leq P\left[\frac{p_i}{p_i + (1-p_i)e^{\varepsilon}} \leq X_i \leq \frac{p_i}{p_i + (1-p_i)e^{-\varepsilon}} \right] \\
       &= P\left[\frac{p_i}{p_i + (1-p_i)e^{\varepsilon}} \leq \frac{p_i}{p_i + (1-p_i)e^{-Z}} \leq \frac{p_i}{p_i + (1-p_i)e^{-\varepsilon}} \right] \\
       &= P[-\varepsilon \leq Z \leq \varepsilon].
   \end{align*}
   Now suppose $\data_2$ has one more row than $\data_1$. Let $i$ index the observation that is in $\data_2$, but not $\data_1$. By Theorem \ref{thm:PLRV_posterior} with $\data = \data_2$, it follows that if $Y \leftarrow M(\data_1)$, then $X_i = p_i/(p_i + (1-p_i)e^{Z})$. Thus, from (\ref{eq:eps_delta_post}) with $Y \leftarrow M(\data_1)$,
   \begin{align*}
       1-\delta &\leq P\left[\frac{p_i}{p_i + (1-p_i)e^{\varepsilon}} \leq X_i \leq \frac{p_i}{p_i + (1-p_i)e^{-\varepsilon}} \right] \\
       &= P\left[\frac{p_i}{p_i + (1-p_i)e^{\varepsilon}} \leq \frac{p_i}{p_i + (1-p_i)e^{Z}} \leq \frac{p_i}{p_i + (1-p_i)e^{-\varepsilon}} \right] \\
       &= P[-\varepsilon \leq Z \leq \varepsilon].
   \end{align*}
   Thus, for any pair of neighboring databases the absolute PLRV is bounded by $\varepsilon$ with probability $1-\delta$ and so $M$ satisfies $(\varepsilon,\delta)$-PDP.
\end{proof}

\subsection{Proof of Theorem \ref{thm:PDP_ratio}} \label{proof:PDP_ratio}

We now prove Theorem \ref{thm:PDP_ratio}, which is restated below.

\begin{reptheorem}{thm:PDP_ratio}
    For the setting described in Section \ref{sec:setting}, let the posterior-to-prior ratio be given by
    \begin{align} \tag{\ref{eq:r}}
        r_i(y, p_i) = \frac{P_{\M}[I_i = 1 \mid Y = y]}{P_{\M}[I_i = 1]}.
    \end{align}
    Let $R_i = r_i(Y, p_i)$. Then, $M$ satisfies $(\varepsilon, \delta)$-PDP if and only if for any database $\data$, target $i$, $p_i \in (0,1]$, and under both $Y \leftarrow M(\data)$ and $Y \leftarrow M(\data_{-i})$,
    \begin{align} \tag{\ref{eq:eps_delta_ratio}}
        P\left[e^{-\varepsilon} \leq R_i \leq e^{\varepsilon} \right] \geq 1-\delta.
    \end{align}
\end{reptheorem}

\begin{proof}
    ($\Longrightarrow$) By Theorem \ref{thm:PDP_posterior}, $(\varepsilon, \delta)$-probabilistic differential privacy implies that in the setting described in Section \ref{sec:setting}, for any database $\data$, target $i$, $p_i \in (0,1]$, and under both $Y \leftarrow M(\data)$ and $Y \leftarrow M(\data_{-i})$, the distribution of the adversary's posterior probability, $X_i$, satisfies
    \begin{align*}
        1-\delta &\leq P\left[\frac{p_i}{p_i + (1-p_i)e^{\varepsilon}} \leq X_i \leq \frac{p_i}{p_i + (1-p_i)e^{-\varepsilon}} \right] \\
        &= P\left[\frac{1}{p_i + (1-p_i)e^{\varepsilon}} \leq \frac{X_i}{p_i} \leq \frac{1}{p_i + (1-p_i)e^{-\varepsilon}} \right].
    \end{align*}
    Letting $f_i(y, p_i)$ be defined as in (\ref{eq:f}), it follows that $r_i(y, p_i) = f_i(y, p_i)/p_i$ and so $R_i = f_i(Y, p_i)/ p_i = X_i/p_i$. Since $p_i(1-e^{-\varepsilon}) \geq 0$, it follows that
    \begin{align*}
        \frac{1}{p_i + (1-p_i)e^{-\varepsilon}} = \frac{1}{e^{-\varepsilon} + p_i(1-e^{-\varepsilon})} \leq \frac{1}{e^{-\varepsilon}} = e^\varepsilon
    \end{align*}
    and since $p_i(e^{\varepsilon}-1) \geq 0$, it follows that
    \begin{align*}
        \frac{1}{p_i + (1-p_i)e^{\varepsilon}} = \frac{1}{e^{\varepsilon} - p_i(e^{\varepsilon} - 1)} \geq \frac{1}{e^{\varepsilon}} = e^{-\varepsilon}.
    \end{align*}
    Thus, if $M$ satisfies $(\varepsilon,\delta)$-probabilistic differential privacy, then for any database $\data$, target $i$, $p_i \in (0,1]$, and under both $Y \leftarrow M(\data)$ and $Y \leftarrow M(\data_{-i})$,
    \begin{align*}
        P\left[e^{-\varepsilon} \leq R_i \leq e^\varepsilon \right] &\geq P\left[\frac{1}{p_i + (1-p_i)e^{\varepsilon}} \leq R_i \leq \frac{1}{p_i + (1-p_i)e^{-\varepsilon}} \right] \\
        &\geq 1-\delta.
    \end{align*}
    The result follows.

    ($\Longleftarrow$) Now suppose (\ref{eq:eps_delta_ratio}) is satisfied for all $\data$, all $i$, and all $p_i \in (0,1]$, under both $Y \leftarrow M(\data)$ and $Y \leftarrow M(\data_{-i})$. Let $\data_1, \data_2$ be an arbitrary pair of neighboring databases and let $Z = \PL(M(\data_1) ~\Vert~ M(\data_2))$.

    Suppose $\data_1$ has one more row than $\data_2$ and let $i$ index the observation that is in $\data_1$, but not $\data_2$. By Theorem \ref{thm:PLRV_posterior} with $\data = \data_1$, it follows that if $Y \leftarrow M(\data_1)$, then $X_i = p_i/(p_i + (1-p_i)e^{-Z})$. As noted above, $R_i = X_i/p_i$. Thus, from (\ref{eq:eps_delta_ratio}) with $Y \leftarrow M(\data_1)$, for all $p_i \in (0,1]$,
    \begin{align*}
        1-\delta &\leq P\left[e^{-\varepsilon} \leq \frac{X_i}{p_i} \leq e^\varepsilon \right] \\
        &= P\left[e^{-\varepsilon} \leq \frac{1}{p_i + (1-p_i)e^{-Z}} \leq e^\varepsilon \right] \\
        &= P\left[e^{-\varepsilon} \leq p_i + (1-p_i)e^{-Z} \leq e^\varepsilon \right].
    \end{align*}
    Since this holds for all $p_i \in (0,1]$, it also holds in the limit as $p_i \to 0$. This implies
    \begin{align*}
        1-\delta &\leq P\left[e^{-\varepsilon} \leq e^{-Z} \leq e^\varepsilon \right] = P\left[-\varepsilon \leq Z \leq \varepsilon \right]. 
    \end{align*}
    
    Now suppose $\data_2$ has one more row than $\data_1$. Let $i$ index the observation that is in $\data_2$, but not $\data_1$. By Theorem \ref{thm:PLRV_posterior} with $\data = \data_2$, it follows that if $Y \leftarrow M(\data_1)$, then $X_i = p_i/(p_i + (1-p_i)e^{Z})$. Thus, from (\ref{eq:eps_delta_post}) with $Y \leftarrow M(\data_1)$, for all $p_i \in (0,1]$,
    \begin{align*}
        1-\delta &\leq P\left[e^{-\varepsilon} \leq \frac{X_i}{p_i} \leq e^\varepsilon \right] \\
        &= P\left[e^{-\varepsilon} \leq \frac{1}{p_i + (1-p_i)e^{Z}} \leq e^\varepsilon \right] \\
        &= P\left[e^{-\varepsilon} \leq p_i + (1-p_i)e^{Z} \leq e^\varepsilon \right].
    \end{align*}
    Since this holds for all $p_i \in (0,1]$, it also holds in the limit as $p_i \to 0$. This implies
    \begin{align*}
        1-\delta &\leq P\left[e^{-\varepsilon} \leq e^{Z} \leq e^\varepsilon \right] = P\left[-\varepsilon \leq Z \leq \varepsilon \right]. 
    \end{align*}
    Thus, for any pair of neighboring databases the absolute privacy-loss is bounded by $\varepsilon$ with probability $1-\delta$. Thus, $M$ satisfies $(\varepsilon,\delta)$-probabilistic DP.
\end{proof}

\subsection{Proof of Lemma \ref{lem:diff_worst}} \label{proof:diff_worst}

We now prove Lemma \ref{lem:diff_worst}, which is restated below.

\begin{replemma}{lem:diff_worst}
    Suppose that for all $p_i \in [0,1]$, a random variable $D_i$ is such that
    \begin{align} \tag{\ref{eq:D_i_bounds}}
        P\left[\frac{p_i}{p_i + (1-p_i)e^{\varepsilon}} - p_i \leq D_i \leq \frac{p_i}{p_i + (1-p_i)e^{-\varepsilon}} - p_i \right] \geq 1-\delta.
    \end{align}
    Then,
    \begin{enumerate}
        \item The lower limit of the interval in (\ref{eq:D_i_bounds}) is minimized when $p_i = 1/ (1+e^{-\varepsilon/2})$.
    
        \item The upper limit of the interval in (\ref{eq:D_i_bounds}) is maximized when $p_i = 1/(1+e^{\varepsilon/2})$.
    \end{enumerate}
\end{replemma}

\begin{proof}
    Define $g(p_i, a) = p_i/(p_i + (1-p_i)a) - p_i$ for $a > 0$. We can then rewrite (\ref{eq:D_i_bounds}) as
\begin{align*}
    P[g(p_i, e^{\varepsilon}) \leq D_i \leq g(p_i, e^{-\varepsilon})] \geq 1-\delta.
\end{align*}
Differentiating $g$ with respect to $p_i$ yields
\begin{align*}
    \frac{\partial g(p_i, a)}{\partial p_i} &= \frac{p_i + (1-p_i)a - p_i(1-a)}{(p_i + (1-p_i)a)^2} - 1 \\
    &= \frac{a}{(p_i + (1-p_i)a)^2} - 1.
\end{align*}
Setting the derivative equal to zero and solving yields the following stationary point.
\begin{align*}
    \frac{\partial g(p_i,a)}{\partial p_i} = 0 \implies \sqrt{a} = p_i + (1-p_i)a \implies p_i = \frac{\sqrt{a}}{1 + \sqrt{a}}.
\end{align*}
Taking the second derivative yields
\begin{align*}
    \frac{\partial^2 g(p_i,a)}{\partial p_i^2}  = -\frac{2a(1-a)}{(p_i + (1-p_i)a)^3}.
\end{align*}
It is straightforward to verify that at the stationary point, the second derivative is negative when $a < 1$ and positive when $a > 1$. Thus, $g(p_i, e^{\varepsilon})$ is minimized by
\begin{align*}
    \tilde{p}_i' = \frac{e^{\varepsilon/2}}{1+e^{\varepsilon/2}} = \frac{1}{e^{-\varepsilon/2} + 1},
\end{align*}
yielding part 1 of the result. Similarly, $g(p_i, e^{-\varepsilon})$ is maximized by
\begin{align*}
    \tilde{p}_i = \frac{e^{-\varepsilon/2}}{1+e^{-\varepsilon/2}} = \frac{1}{e^{\varepsilon/2} + 1},
\end{align*}
yielding part 2 of the result.
\end{proof}

\subsection{Proof of Theorem \ref{thm:PDP_diff}} \label{proof:PDP_diff}

We now prove Theorem \ref{thm:PDP_diff}, which is restated below.

\begin{reptheorem}{thm:PDP_diff}
    For the setting described in Section \ref{sec:setting}, let the posterior-to-prior difference be given by
    \begin{align} \tag{\ref{eq:d}}
        d_i(y, p_i) = P_{\M}[I_i = 1 \mid Y = y] - P_{\M}[I_i = 1].
    \end{align}
    Let $D_i = d_i(Y, p_i)$. Then,
    \begin{enumerate}
        \item If $M$ satisfies $(\varepsilon, \delta)$-PDP, then for any database $\data$, target $i$, $p_i \in [0,1]$, and under both $Y \leftarrow M(\data)$ and $Y \leftarrow M(\data_{-i})$,
        \begin{align} \tag{\ref{eq:eps_delta_diff}}
            P\left[-\frac{e^{\varepsilon/2} - 1}{e^{\varepsilon/2} + 1} \leq D_i \leq \frac{e^{\varepsilon/2} - 1}{e^{\varepsilon/2} + 1} \right] \geq 1-\delta.
        \end{align}

        \item If $M$ is such that for any database $\data$, target $i$, $p_i \in [0,1]$, and under both $Y \leftarrow M(\data)$ and $Y \leftarrow M(\data_{-i})$,
        \begin{align} \tag{\ref{eq:eps_delta_diff2}}
            P\left[-\frac{e^{\varepsilon/2} - 1}{e^{\varepsilon/2} + 1} \leq D_i \leq \frac{e^{\varepsilon/2} - 1}{e^{\varepsilon/2} + 1} \right] \geq 1-\delta,
        \end{align}
        then $M$ satisfies both $(\varepsilon, 2\delta)$-PDP and $(\tilde{\varepsilon}, \delta)$-PDP for $\tilde{\varepsilon} = \log(3e^{\varepsilon/2} - 1) - \log(3 - e^{\varepsilon/2})$.
    \end{enumerate}
\end{reptheorem}

\begin{proof}
    (Part 1) 
    By Theorem \ref{thm:PDP_posterior}, $(\varepsilon, \delta)$-probabilistic differential privacy implies that for any database $\data$, in the setting described in Section \ref{sec:setting}, for any target $i$, $p_i \in [0,1]$, and under both $Y \leftarrow M(\data)$ and $Y \leftarrow M(\data_{-i})$, the distribution of the adversary's posterior probability, $X_i$, satisfies
    \begin{align}
        &1-\delta \nonumber \\
        &\leq P\left[\frac{p_i}{p_i + (1-p_i)e^{\varepsilon}} \leq X_i \leq \frac{p_i}{p_i + (1-p_i)e^{-\varepsilon}} \right] \nonumber \\
        &= P\left[\frac{p_i}{p_i + (1-p_i)e^{\varepsilon}} - p_i \leq X_i - p_i \leq \frac{p_i}{p_i + (1-p_i)e^{-\varepsilon}} - p_i \right]. \label{eq:D_bounds}
    \end{align}
    
    Letting $f_i(y, p_i)$ be defined as in (\ref{eq:f}), it follows that $d_i(y, p_i) = f_i(y, p_i) - p_i$ and so $D_i = f_i(Y, p_i) - p_i = X_i - p_i$. By Lemma \ref{lem:diff_worst}, the lower bound in (\ref{eq:D_bounds}) is minimized by $\tilde{p}_i' = 1/(e^{-\varepsilon/2} + 1)$ and so the function takes minimum value
    \begin{align*}
        \frac{\tilde{p}_i'}{\tilde{p}_i' + \left(1 - \tilde{p}_i'\right)e^{\varepsilon}} - \tilde{p}_i' = \frac{1}{1 + e^{-\varepsilon/2}e^{\varepsilon}} - \frac{1}{1+e^{-\varepsilon/2}} = -\frac{e^{\varepsilon/2} - 1}{e^{\varepsilon/2} + 1}.
    \end{align*}
    Similarly, by Lemma \ref{lem:diff_worst} the upper bound in (\ref{eq:D_bounds}) is maximized by $\tilde{p}_i = 1/(e^{\varepsilon/2} + 1)$ and so the function takes maximum value
    \begin{align*}
        \frac{\tilde{p}_i}{\tilde{p}_i + \left(1 - \tilde{p}_i\right)e^{-\varepsilon}} - \tilde{p}_i = \frac{1}{1 + e^{\varepsilon/2}e^{-\varepsilon}} - \frac{1}{1+e^{\varepsilon/2}} = \frac{e^{\varepsilon/2} - 1}{e^{\varepsilon/2} + 1}.
    \end{align*}
    
    Thus, for any $p_i$, 
    \begin{align*}
        &P\left[-\frac{e^{\varepsilon/2} - 1}{e^{\varepsilon/2} + 1} \leq D_i \leq \frac{e^{\varepsilon/2} - 1}{e^{\varepsilon/2} + 1} \right] \\
        &\geq  P\left[\frac{p_i}{p_i + (1-p_i)e^{\varepsilon}} - p_i \leq D_i \leq \frac{p_i}{p_i + (1-p_i)e^{-\varepsilon}} - p_i \right] \\
        &\geq 1-\delta.
    \end{align*}

    (Part 2) Suppose (\ref{eq:eps_delta_diff2}) is satisfied for all $\data$, all $i$, and all $p_i \in [0,1]$, under both $Y \leftarrow M(\data)$ and $Y \leftarrow M(\data_{-i})$. Let $\data_1, \data_2$ be an arbitrary pair of neighboring databases and let $Z = \PL(M(\data_1) ~\Vert~ M(\data_2))$.

    Suppose $\data_1$ has one more row than $\data_2$ and let $i$ index the observation that is in $\data_1$, but not $\data_2$. By Theorem \ref{thm:PLRV_posterior} with $\data = \data_1$, it follows that if $Y \leftarrow M(\data_1)$, then $X_i = p_i/(p_i + (1-p_i)e^{-Z})$. As noted above, $D_i = X_i - p_i$. Thus, from (\ref{eq:eps_delta_diff2}) with $Y \leftarrow M(\data_1)$, for all $p_i \in [0,1]$,
    \begin{align*}
        1-\delta &\leq P\left[-\frac{e^{\varepsilon/2} - 1}{e^{\varepsilon/2} + 1} \leq \frac{p_i}{p_i + (1-p_i)e^{-Z}} - p_i \leq \frac{e^{\varepsilon/2} - 1}{e^{\varepsilon/2} + 1} \right] \\
        &= P\left[\log\left(\frac{1-p_i}{\frac{p_i}{p_i - \frac{e^{\varepsilon/2} - 1}{e^{\varepsilon/2} + 1}} - p_i}\right) \leq Z \leq \log\left(\frac{1-p_i}{\frac{p_i}{p_i+\frac{e^{\varepsilon/2} - 1}{e^{\varepsilon/2} + 1}} - p_i}\right) \right].
    \end{align*}
    Taking $p_i = 1/2$ and simplifying yields
    \begin{align}
        1-\delta \leq P\left[-\log\left(\frac{3e^{\varepsilon/2} - 1}{3 - e^{\varepsilon/2}}\right) \leq Z \leq \log\left(\frac{3e^{\varepsilon/2} - 1}{3 - e^{\varepsilon/2}}\right) \right], \label{eq:p_half_1}
    \end{align}
    taking $p_i = 1/(1+e^{\varepsilon/2})$ and simplifying yields
    \begin{align}
        1-\delta \leq P\left[-\log\left(\frac{2e^{\varepsilon/2} - 1}{2e^{\varepsilon/2} - e^\varepsilon}\right) \leq Z \leq \varepsilon \right], \label{eq:p_max_1}
    \end{align}
    and taking $p_i = 1/(1+e^{-\varepsilon/2})$ and simplifying yields
    \begin{align}
        1-\delta \leq P\left[-\varepsilon \leq Z \leq \log\left(\frac{2e^{\varepsilon/2} - 1}{2e^{\varepsilon/2} - e^\varepsilon}\right) \right]. \label{eq:p_min_1}
    \end{align}
    
    Now suppose $\data_2$ has one more row than $\data_1$. Let $i$ index the observation that is in $\data_2$, but not $\data_1$. By Theorem \ref{thm:PLRV_posterior} with $\data = \data_2$, it follows that if $Y \leftarrow M(\data_1)$, then $X_i = p_i/(p_i + (1-p_i)e^{Z})$. Thus, from (\ref{eq:eps_delta_post}) with $Y \leftarrow M(\data_1)$, for all $p_i \in (0,1]$,
    \begin{align*}
        1-\delta &\leq P\left[-\frac{e^{\varepsilon/2} - 1}{e^{\varepsilon/2} + 1} \leq \frac{p_i}{p_i + (1-p_i)e^{Z}} - p_i \leq \frac{e^{\varepsilon/2} - 1}{e^{\varepsilon/2} + 1} \right] \\
        &= P\left[-\log\left(\frac{1-p_i}{\frac{p_i}{p_i+\frac{e^{\varepsilon/2} - 1}{e^{\varepsilon/2} + 1}} - p_i}\right) \leq Z \leq -\log\left(\frac{1-p_i}{\frac{p_i}{p_i - \frac{e^{\varepsilon/2} - 1}{e^{\varepsilon/2} + 1}} - p_i}\right)\right].
    \end{align*}
    Taking $p_i = 1/2$ and simplifying yields
    \begin{align}
        1-\delta \leq P\left[-\log\left(\frac{3e^{\varepsilon/2} - 1}{3 - e^{\varepsilon/2}}\right) \leq Z \leq \log\left(\frac{3e^{\varepsilon/2} - 1}{3 - e^{\varepsilon/2}}\right) \right], \label{eq:p_half_2}
    \end{align}
    taking $p_i = 1/(1+e^{\varepsilon/2})$ and simplifying yields
    \begin{align}
        1-\delta \leq P\left[-\varepsilon \leq Z \leq \log\left(\frac{2e^{\varepsilon/2} - 1}{2e^{\varepsilon/2} - e^\varepsilon}\right) \right], \label{eq:p_max_2}
    \end{align}
    and taking $p_i = 1/(1+e^{-\varepsilon/2})$ and simplifying yields
    \begin{align}
        1-\delta \leq P\left[-\log\left(\frac{2e^{\varepsilon/2} - 1}{2e^{\varepsilon/2} - e^\varepsilon}\right) \leq Z \leq \varepsilon \right]. \label{eq:p_min_2}
    \end{align}
    
    From (\ref{eq:p_half_1}) and (\ref{eq:p_half_2}), 
    for any pair of neighboring databases the absolute privacy-loss is bounded by $\tilde{\varepsilon}$ with probability $1-\delta$. Thus, $M$ satisfies $(\tilde{\varepsilon},\delta)$-probabilistic DP.

    From (\ref{eq:p_max_1}) and (\ref{eq:p_min_2}), it follows that for any pair of neighboring databases, $P[Z \geq -\varepsilon] \geq 1-\delta$, and from (\ref{eq:p_min_1}) and (\ref{eq:p_max_2}), it follows that for any pair of neighboring databases, $P[Z \leq \varepsilon] \geq 1-\delta$. Combining these facts yields that for any pair of neighboring databases the absolute privacy-loss is bounded by $\varepsilon$ with probability $1-2\delta$. Thus, $M$ satisfies $(\varepsilon,2\delta)$-probabilistic DP.
\end{proof}

\subsection{Mechanism with Bounded Posterior-to-Prior Difference That Does Not Satisfy PDP} \label{proof:counterexample}

It may be surprising that a mechanism satisfying
\begin{align*}
    P\left[-\frac{e^{\varepsilon/2} - 1}{e^{\varepsilon/2} + 1} \leq D_i \leq \frac{e^{\varepsilon/2} - 1}{e^{\varepsilon/2} + 1} \right] \geq 1-\delta
\end{align*}
for all $p_i \in [0,1]$ need not satisfy $(\varepsilon, \delta)$-probabilistic DP. We provide the following counterexample. 

\begin{example}
    Let $\varepsilon,\delta > 0$ and let $\tilde{\varepsilon} = \log(3e^{\varepsilon/2} - 1) - \log(3 - e^{\varepsilon/2})$. Define $M$ to be a mechanism that performs randomized response under $\tilde{\varepsilon}$-DP with probability $\delta(1 + e^{-\tilde{\varepsilon}})$ and releases nothing otherwise.
\end{example}

By a straightforward computation, an adversary's posterior-to-prior difference takes the form
\begin{align*}
    D_i = \begin{cases}
        0, & \mbox{with probability } 1-\delta(1+e^{-\tilde{\varepsilon}}); \\
        \frac{p_i}{p_i + (1-p_i)e^{-\tilde{\varepsilon}}} - p_i, & \mbox{with probability } \delta; \\
        \frac{p_i}{p_i + (1-p_i)e^{\tilde{\varepsilon}}} - p_i, & \mbox{with probability } \delta e^{-\tilde{\varepsilon}}.
    \end{cases}
\end{align*}
When $p_i \geq 1/2$, it can be verified that
\begin{align*}
    \frac{p_i}{p_i + (1-p_i)e^{-\tilde{\varepsilon}}} - p_i\leq \frac{e^{\varepsilon/2} - 1}{e^{\varepsilon/2} + 1}.
\end{align*}
Thus, for $p_i \geq 1/2$, we have that $D_i \leq (e^{\varepsilon/2} - 1)/(e^{\varepsilon/2} + 1)$ with probability at least $1-\delta(1+e^{-\tilde{\varepsilon}}) + \delta > 1-\delta$. Similarly, when $p_i \leq 1/2$, it can be verified that
\begin{align*}
    \frac{p_i}{p_i + (1-p_i)e^{\tilde{\varepsilon}}} - p_i \geq -\frac{e^{\varepsilon/2} - 1}{e^{\varepsilon/2} + 1}.
\end{align*}
Thus, for $p_i \leq 1/2$, we have that $D_i \geq -(e^{\varepsilon/2} - 1)/(e^{\varepsilon/2} + 1)$ with probability at least $1-\delta(1+e^{-\tilde{\varepsilon}}) + \delta e^{-\tilde{\varepsilon}} = 1-\delta$. Hence, the boundedness condition for the posterior-to-prior difference is satisfied, but $P[-\varepsilon \leq Z \leq \varepsilon] = 1 - \delta(1+e^{-\tilde{\varepsilon}}) < 1-\delta$ and $M$ does not satisfy $(\varepsilon, \delta)$-probabilistic DP for any $\varepsilon > 0$ and $\delta > 0$.

\subsection{Proofs of Corollaries \ref{cor:pure_DP_posterior}-\ref{cor:pure_DP_diff}} \label{proof:pure_DP} 

We now prove the corollaries relating pure differential privacy to disclosure risk metrics. The results are each restated below.

\begin{repcorollary}{cor:pure_DP_posterior}
    $M$ satisfies $(\varepsilon, 0)$-DP if and only if in the setting described in Section \ref{sec:setting}, for any database $\data$, target $i$, $p_i \in [0,1]$, and under both $Y \leftarrow M(\data)$ and $Y \leftarrow M(\data_{-i})$, the distribution of $X_i$ satisfies
    \begin{align} \tag{\ref{eq:eps_post}}
        P\left[\frac{p_i}{p_i + (1-p_i)e^{\varepsilon}} \leq X_i \leq \frac{p_i}{p_i + (1-p_i)e^{-\varepsilon}} \right] = 1.
    \end{align}
\end{repcorollary}

\begin{proof}
    From Theorem \ref{thm:pure_DP_PLRV}, $M$ satisfies $(\varepsilon, 0)$-DP if and only if $M$ satisfies $(\varepsilon, 0)$-PDP. From Theorem \ref{thm:PDP_posterior}, $M$ satisfies $(\varepsilon, 0)$-PDP if and only if for any database $\data$, in the setting described in Section \ref{sec:setting}, for any target $i$, $p_i \in [0,1]$, and under both $Y \leftarrow M(\data)$ and $Y \leftarrow M(\data_{-i})$, the distribution of the adversary's posterior probability, $X_i$, satisfies
    \begin{align} \label{eq:eps_post_geq}
        P\left[\frac{p_i}{p_i + (1-p_i)e^{\varepsilon}} \leq X_i \leq \frac{p_i}{p_i + (1-p_i)e^{-\varepsilon}} \right] \geq 1.
    \end{align}
    Since the quantity in (\ref{eq:eps_post_geq}) is a probability, it is bounded above by 1. The result follows.
\end{proof}

\begin{repcorollary}{cor:pure_DP_ratio}
    $M$ satisfies $(\varepsilon,0)$-DP if and only if in the setting described in Section \ref{sec:setting}, for any database $\data$, target $i$, $p_i \in (0,1]$, and under both $Y \leftarrow M(\data)$ and $Y \leftarrow M(\data_{-i})$, the distribution of $R_i$, as defined in Theorem \ref{thm:PDP_ratio}, satisfies
    \begin{align} \tag{\ref{eq:pure_DP_ratio}}
        P\left[e^{-\varepsilon} \leq R_i \leq e^{\varepsilon} \right] = 1.
    \end{align}
\end{repcorollary}

\begin{proof}
    From Theorem \ref{thm:pure_DP_PLRV}, $M$ satisfies $(\varepsilon, 0)$-DP if and only if $M$ satisfies $(\varepsilon, 0)$-PDP. From Theorem \ref{thm:PDP_ratio}, $M$ satisfies $(\varepsilon, \delta)$-PDP if and only if for any database $\data$, target $i$, $p_i \in (0,1]$, and under both $Y \leftarrow M(\data)$ and $Y \leftarrow M(\data_{-i})$,
    \begin{align} \label{eq:eps_ratio_geq}
        P\left[e^{-\varepsilon} \leq R_i \leq e^{\varepsilon} \right] \geq 1.
    \end{align}
    Since the quantity in (\ref{eq:eps_ratio_geq}) is a probability, it is bounded above by 1. The result follows.
\end{proof}

\begin{repcorollary}{cor:pure_DP_diff}
    $M$ satisfies $(\varepsilon,0)$-DP if and only if in the setting described in Section \ref{sec:setting}, for any database $\data$, target $i$, $p_i \in [0,1]$, and under both $Y \leftarrow M(\data)$ and $Y \leftarrow M(\data_{-i})$, the distribution of $D_i$, as defined in Theorem \ref{thm:PDP_diff}, satisfies
    \begin{align} \tag{\ref{eq:pure_DP_diff}}
        P\left[-\frac{e^{\varepsilon/2} - 1}{e^{\varepsilon/2} + 1} \leq D_i \leq \frac{e^{\varepsilon/2} - 1}{e^{\varepsilon/2} + 1} \right] = 1.
    \end{align}
\end{repcorollary}

\begin{proof}
    From Theorem \ref{thm:pure_DP_PLRV}, $M$ satisfies $(\varepsilon, 0)$-DP if and only if $M$ satisfies $(\varepsilon, 0)$-PDP. From Theorem \ref{thm:PDP_diff}, $M$ satisfies $(\varepsilon, \delta)$-PDP if and only if for any database $\data$, target $i$, $p_i \in (0,1]$, and under both $Y \leftarrow M(\data)$ and $Y \leftarrow M(\data_{-i})$,
    \begin{align} \label{eq:eps_diff_geq}
        P\left[-\frac{e^{\varepsilon/2} - 1}{e^{\varepsilon/2} + 1} \leq D_i \leq \frac{e^{\varepsilon/2} - 1}{e^{\varepsilon/2} + 1} \right] \geq 1.
    \end{align}
    Since the quantity in (\ref{eq:eps_diff_geq}) is a probability, it is bounded above by 1. The result follows.
\end{proof}

\subsection{Proof of Theorem \ref{thm:DP_to_PDP}} \label{proof:DP_to_PDP}

We now prove Theorem \ref{thm:DP_to_PDP}. We note that authors of \cite{zhao2019reviewing} demonstrate that $(\varepsilon, \delta)$-DP implies $(\varepsilon', \delta')$-PDP for all $\varepsilon' > \varepsilon$ and $\delta' = \delta(1+e^{-\varepsilon'})/(1-e^{\varepsilon-\varepsilon'})$. It can be shown that Theorem \ref{thm:DP_to_PDP} 
is implied by this result. However, the proof in \cite{zhao2019reviewing} is fairly technical. We will provide a direct, less technical proof of Theorem \ref{thm:DP_to_PDP} that does not utilize the result in \cite{zhao2019reviewing}. For this proof, we leverage the following characterization of approximate DP in terms of privacy loss random variables.

\begin{theorem} \label{thm:ADP_PLRV}
    \cite{canonne2020discrete}
    Let $M$ be a mechanism and let $\data$ and $\data'$ be neighboring databases. Define $Z = \PL(M(\data) \,\Vert\, M(\data'))$ and $Z' = \PL(M(\data') \,\Vert\, M(\data))$. Then $M$ satisfies $(\varepsilon, \delta)$-DP if and only if
    \begin{align}
        \delta &\geq P[Z > \varepsilon] - e^{\varepsilon}P[-Z' > \varepsilon] \\
        &= E[\max\{0, 1-e^{\varepsilon-Z}\}] \\
        &= \int_\varepsilon^\infty e^{\varepsilon - z} P[Z > z] \, \mathrm{d}z.
    \end{align}
\end{theorem}

The result to be proved is restated below.

\begin{reptheorem}{thm:DP_to_PDP}
    If $M$ satisfies $(\varepsilon, \delta)$-DP, then for any $\delta' \in (\delta, 1]$, $M$ satisfies $(\varepsilon', \delta')$-PDP for $\varepsilon' = \log(\delta'e^{\varepsilon} +\delta) - \log(\delta' - \delta)$.
\end{reptheorem}

\begin{proof}
    Let $\data$ and $\data'$ be neighboring databases and define $Z = \PL(M(\data) \,\Vert\, M(\data'))$ and $Z' = \PL(M(\data') \,\Vert\, M(\data))$. Let $\delta' \in (\delta, 1]$ and $\varepsilon' = \log(\delta'e^{\varepsilon} +\delta) - \log(\delta' - \delta)$.  Note that $\varepsilon' > \varepsilon$, since 
    \begin{align*}
        \varepsilon' &= \log\left(e^\varepsilon\frac{\delta' + \delta e^{-\varepsilon}}{\delta' - \delta} \right) \\
        &= \varepsilon + \log\left(\frac{\delta' - \delta + \delta + \delta e^{-\varepsilon}}{\delta' - \delta} \right) \\
        &= \varepsilon + \log\left(1 + \delta \, \frac{1 + e^{-\varepsilon}}{\delta' - \delta} \right) \\
        &> \varepsilon.
    \end{align*}
    Then,
    \begin{align}
        P[-\varepsilon' \leq Z \leq \varepsilon'] &= 1 - P[Z > \varepsilon'] - P[-Z > \varepsilon'] \nonumber \\
        &= 1 - P[Z > \varepsilon'] \nonumber \\
        &\qquad + e^{-\varepsilon'}\left(-e^{\varepsilon'}P[-Z > \varepsilon'] + P[Z' > \varepsilon']\right) \nonumber \\
        &\qquad -  e^{-\varepsilon'}P[Z' > \varepsilon']. \label{eq:PDP_parentheses}
    \end{align}
    Recall that since $M$ satisfies $(\varepsilon, \delta)$-DP and $\varepsilon' > \varepsilon$, it follows that $M$ satisfies $(\varepsilon', \delta)$-DP. Then, by Theorem \ref{thm:ADP_PLRV}, the quantity in parentheses in (\ref{eq:PDP_parentheses}) is nonnegative:
    \begin{align*}
        -e^{\varepsilon'}P[-Z > \varepsilon'] + P[Z' > \varepsilon'] &= E[\max\{0, 1-e^{\varepsilon'-Z'}\}] \\ 
        &\geq  E[0] \\
        &= 0.
    \end{align*}
    Thus,
    \begin{align*}
        P[-\varepsilon' \leq Z \leq \varepsilon']
        &\geq 1 - P[Z > \varepsilon'] -  e^{-\varepsilon'}P[Z' > \varepsilon'].
    \end{align*}
    Applying Markov's inequality to each term with the function $f(x) = \max\{0, 1-e^{\varepsilon - x}\}$ and then bounding the numerator via Theorem \ref{thm:ADP_PLRV} gives
    \begin{align*}
        &P[-\varepsilon' \leq Z \leq \varepsilon'] \\
        &\geq 1 - \frac{E[\max\{0, 1-e^{\varepsilon - Z}\}]}{1 - e^{\varepsilon - \varepsilon'}} -  e^{-\varepsilon'} \frac{E[\max\{0, 1-e^{\varepsilon - Z'}\}]}{1 - e^{\varepsilon - \varepsilon'}} \\
        &\geq 1 - \frac{\delta}{1 - e^{\varepsilon - \varepsilon'}} -  e^{-\varepsilon'} \frac{\delta}{1 - e^{\varepsilon - \varepsilon'}}.
    \end{align*}
    A straightforward computation yields that $e^{-\varepsilon'} = (\delta' - \delta)/(\delta' e^\varepsilon + \delta)$. Thus,
    \begin{align*}
        P[-\varepsilon' \leq Z \leq \varepsilon'] 
        &\geq 1 - \delta\frac{1 + e^{-\varepsilon'}}{1 - e^{\varepsilon - \varepsilon'}} \\
        &= 1- \delta \frac{1 + \frac{\delta' - \delta}{\delta' e^\varepsilon + \delta}}{1 - e^{\varepsilon}\frac{\delta' - \delta}{\delta' e^\varepsilon + \delta}} \\
        &= 1- \delta \frac{\delta' e^\varepsilon + \delta + \delta' - \delta}{\delta' e^\varepsilon + \delta - e^{\varepsilon}(\delta' - \delta)} \\
        &= 1-\delta \frac{\delta'(1 + e^\varepsilon)}{\delta(1 + e^\varepsilon)} \\
        &=1- \delta'.
    \end{align*}
    The result follows.
\end{proof}

\subsection{Proof of Corollaries \ref{cor:approx_DP_posterior}-\ref{cor:approx_DP_diff}} \label{proof:approx_DP}

We now prove the corollaries relating approximate differential privacy to disclosure risk metrics. The results are each restated below.

\begin{repcorollary}{cor:approx_DP_posterior}
    In the setting described in Section \ref{sec:setting},
    \begin{enumerate}
        \item If $M$ is such that for any database $\data$, target $i$, $p_i \in [0,1]$, and under both $Y \leftarrow M(\data)$ and $Y \leftarrow M(\data_{-i})$, the distribution of $X_i$ satisfies
        \begin{align} \tag{\ref{eq:post_to_eps_delta}}
            P\left[\frac{p_i}{p_i + (1-p_i)e^{\varepsilon}} \leq X_i \leq \frac{p_i}{p_i + (1-p_i)e^{-\varepsilon}} \right] \geq 1-\delta,
        \end{align}
        then $M$ satisfies $(\varepsilon, \delta)$-DP.

        \item If $M$ satisfies $(\varepsilon, \delta)$-DP, then for any $\delta' > \delta$, database $\data$, target $i$, $p_i \in [0,1]$, and under both $Y \leftarrow M(\data)$ and $Y \leftarrow M(\data_{-i})$, for $\varepsilon' = \log(\delta'e^{\varepsilon} +\delta) - \log(\delta' - \delta)$, the distribution of $X_i$ satisfies
        \begin{align} \tag{\ref{eq:eps_delta_to_post}}
            P\left[\frac{p_i}{p_i + (1-p_i)e^{\varepsilon'}} \leq X_i \leq \frac{p_i}{p_i + (1-p_i)e^{-\varepsilon'}} \right] \geq 1-\delta'.
        \end{align}
    \end{enumerate}
\end{repcorollary}

\begin{proof}
    (Part 1) From Theorem \ref{thm:PDP_posterior}, if $M$ is such that (\ref{eq:post_to_eps_delta}) is satisfied for any database $\data$, target $i$, $p_i \in [0,1]$, and under both $Y \leftarrow M(\data)$ and $Y \leftarrow M(\data_{-i})$, then $M$ satisfies $(\varepsilon, \delta)$-PDP. From Theorem \ref{thm:PDP_to_DP}, if $M$ satisfies $(\varepsilon, \delta)$-PDP, then $M$ satisfies $(\varepsilon, \delta)$-DP.

    (Part 2) From Theorem \ref{thm:DP_to_PDP}, if $M$ satisfies $(\varepsilon, \delta)$-DP, then for any $\delta' \in (\delta, 1]$, $M$ satisfies $(\varepsilon', \delta')$-PDP for $\varepsilon' = \log(\delta'e^{\varepsilon} +\delta) - \log(\delta' - \delta)$. From Theorem \ref{thm:PDP_posterior}, if $M$ satisfies $(\varepsilon', \delta')$-PDP, then for any database $\data$, target $i$, $p_i \in [0,1]$, and under both $Y \leftarrow M(\data)$ and $Y \leftarrow M(\data_{-i})$, (\ref{eq:eps_delta_to_post}) is satisfied.
\end{proof}

\begin{repcorollary}{cor:approx_DP_ratio}
    In the setting described in Section \ref{sec:setting}, let $R_i$ be as defined in Theorem \ref{thm:PDP_ratio}.
    \begin{enumerate}
        \item If $M$ is such that for any database $\data$, target $i$, $p_i \in (0,1]$, and under both $Y \leftarrow M(\data)$ and $Y \leftarrow M(\data_{-i})$, $R_i$ satisfies
        \begin{align} \tag{\ref{eq:ratio_to_eps_delta}}
            P\left[e^{-\varepsilon} \leq R_i \leq e^{\varepsilon} \right] \geq 1-\delta,
        \end{align}
        then $M$ satisfies $(\varepsilon, \delta)$-DP.

        \item If $M$ satisfies $(\varepsilon, \delta)$-DP, then for any $\delta' > \delta$, database $\data$, target $i$, $p_i \in (0,1]$, and under both $Y \leftarrow M(\data)$ and $Y \leftarrow M(\data_{-i})$, for $\varepsilon' = \log(\delta'e^{\varepsilon} +\delta) - \log(\delta' - \delta)$,
        \begin{align} \tag{\ref{eq:eps_delta_to_ratio}}
            P\left[e^{-\varepsilon'} \leq R_i \leq e^{\varepsilon'} \right] \geq 1-\delta'.
        \end{align}
    \end{enumerate}
\end{repcorollary}

\begin{proof}
    (Part 1) From Theorem \ref{thm:PDP_ratio}, if $M$ is such that (\ref{eq:ratio_to_eps_delta}) is satisfied for any database $\data$, target $i$, $p_i \in [0,1]$, and under both $Y \leftarrow M(\data)$ and $Y \leftarrow M(\data_{-i})$, then $M$ satisfies $(\varepsilon, \delta)$-PDP. From Theorem \ref{thm:PDP_to_DP}, if $M$ satisfies $(\varepsilon, \delta)$-PDP, then $M$ satisfies $(\varepsilon, \delta)$-DP.

    (Part 2) From Theorem \ref{thm:DP_to_PDP}, if $M$ satisfies $(\varepsilon, \delta)$-DP, then for any $\delta' \in (\delta, 1]$, $M$ satisfies $(\varepsilon', \delta')$-PDP for $\varepsilon' = \log(\delta'e^{\varepsilon} +\delta) - \log(\delta' - \delta)$. From Theorem \ref{thm:PDP_ratio}, if $M$ satisfies $(\varepsilon', \delta')$-PDP, then for any database $\data$, target $i$, $p_i \in [0,1]$, and under both $Y \leftarrow M(\data)$ and $Y \leftarrow M(\data_{-i})$, (\ref{eq:eps_delta_to_ratio}) is satisfied.
\end{proof}

\begin{repcorollary}{cor:approx_DP_diff}
    In the setting described in Section \ref{sec:setting}, let $D_i$ be as defined in Theorem \ref{thm:PDP_diff}.
    \begin{enumerate}
        \item If $M$ is such that for any database $\data$, target $i$, $p_i \in [0,1]$, and under both $Y \leftarrow M(\data)$ and $Y \leftarrow M(\data_{-i})$, the distribution of $D_i$ satisfies
        \begin{align} \tag{\ref{eq:diff_to_eps_delta}}
            P\left[-\frac{e^{\varepsilon/2} - 1}{e^{\varepsilon/2} + 1} \leq D_i \leq \frac{e^{\varepsilon/2} - 1}{e^{\varepsilon/2} + 1} \right] \geq 1-\delta,
        \end{align}
        then $M$ satisfies both $(\varepsilon, 2\delta)$-DP and $(\tilde{\varepsilon}, \delta)$-DP for
        \begin{align*}
            \tilde{\varepsilon} = \log(3e^{\varepsilon/2} - 1) - \log(3 - e^{\varepsilon/2}).
        \end{align*}

        \item If $M$ satisfies $(\varepsilon, \delta)$-DP, then for any $\delta' > \delta$, database $\data$, target $i$, $p_i \in [0,1]$, and under both $Y \leftarrow M(\data)$ and $Y \leftarrow M(\data_{-i})$, for $\varepsilon' = \log(\delta'e^{\varepsilon} +\delta) - \log(\delta' - \delta)$,
        \begin{align} \tag{\ref{eq:eps_delta_to_diff}}
            P\left[-\frac{e^{\varepsilon'/2} - 1}{e^{\varepsilon'/2} + 1} \leq D_i \leq \frac{e^{\varepsilon'/2} - 1}{e^{\varepsilon'/2} + 1} \right] \geq 1-\delta'.
        \end{align}
    \end{enumerate}
\end{repcorollary}

\begin{proof}
    (Part 1) From Theorem \ref{thm:PDP_diff}, if $M$ is such that (\ref{eq:diff_to_eps_delta}) is satisfied for any database $\data$, target $i$, $p_i \in [0,1]$, and under both $Y \leftarrow M(\data)$ and $Y \leftarrow M(\data_{-i})$, then $M$ satisfies $(\varepsilon, \delta)$-PDP. From Theorem \ref{thm:PDP_to_DP}, if $M$ satisfies $(\varepsilon, \delta)$-PDP, then $M$ satisfies both $(\varepsilon, 2\delta)$-probabilistic DP and $(\tilde{\varepsilon}, \delta)$-probabilistic DP for $\tilde{\varepsilon} = \log(3e^{\varepsilon/2} - 1) - \log(3 - e^{\varepsilon/2})$.

    (Part 2) From Theorem \ref{thm:DP_to_PDP}, if $M$ satisfies $(\varepsilon, \delta)$-DP, then for any $\delta' \in (\delta, 1]$, $M$ satisfies $(\varepsilon', \delta')$-PDP for $\varepsilon' = \log(\delta'e^{\varepsilon} +\delta) - \log(\delta' - \delta)$. From Theorem \ref{thm:PDP_diff}, if $M$ satisfies $(\varepsilon', \delta')$-PDP, then for any database $\data$, target $i$, $p_i \in [0,1]$, and under both $Y \leftarrow M(\data)$ and $Y \leftarrow M(\data_{-i})$, (\ref{eq:eps_delta_to_diff}) is satisfied.
\end{proof}

\end{document}